\documentclass[journal]{IEEEtran}

\usepackage{color}
\usepackage{epsf}
\usepackage{times}
\usepackage{epsfig}
\usepackage{graphicx}
\usepackage{url}
\usepackage{amsmath}
\usepackage{amssymb}
\usepackage{amsxtra}
\usepackage[ruled,linesnumbered]{algorithm2e}
\usepackage{cite}
\usepackage{listings}
\usepackage{verbatim}
\DeclareMathOperator*{\argmax}{arg\,max}

\begin{document}
\title{Optimal Denial-of-Service Attack Energy Management over an SINR-Based Network}
\author{Jiahu~Qin,~\IEEEmembership{Senior Member,~IEEE,} Menglin~Li, Ling~Shi,~\IEEEmembership{Senior Member,~IEEE,} \\ and Yu~Kang,~\IEEEmembership{Senior Member,~IEEE}


\thanks{J. Qin  and  M. Li are with the Department of Automation, University of Science and Technology of China, Hefei 230027, China (e-mail: \texttt{jhqin@ustc.edu.cn; lml95@mail.ustc.edu.cn}).}
\thanks{L. Shi is with the Department of Electronic and Computer Engineering, Hong Kong University of Science and Technology, Clear Water Bay, Kowloon, Hong Kong, China (e-mail: \texttt{eesling@ust.hk}).}

\thanks{Y. Kang is with the Department of Automation, the State Key Laboratory of Fire Science, and the Institute of Advanced Technology, University of Science and Technology of China, Hefei 230027, China, and also with the Key Laboratory of Technology in Geo-Spatial Information Processing and Application Systems, Chinese Academy of Sciences, Beijing 100190, China  (e-mail: \texttt{kangduyu@ustc.edu.cn}).}}


\maketitle

\maketitle

\begin{abstract}
We consider a scenario in which a DoS attacker with the limited power resource and the purpose of degrading the system performance, jams a wireless network through which the packet from a sensor is sent to a remote estimator to estimate the system state. To degrade the estimation quality most effectively with a given energy budget, the attacker aims to solve the problem of how much power to obstruct the channel each time, which is the recently proposed optimal attack energy management problem. The existing works are built on an ideal network model in which the packet dropout never occurs when the attack is absent. To encompass wireless transmission losses, we introduce the signal-to-interference-plus-noise ratio-based network. First we focus on the case when the attacker employs the constant power level. To maximize the expected terminal estimation error at the remote estimator, we provide some sufficient conditions for the existence of an explicit solution to the optimal static attack energy management problem and the solution is constructed. Compared with the existing result in which corresponding sufficient conditions work only when the system matrix is normal, the obtained conditions in this paper are viable for a general system and shown to be more relaxed. For the other important index of system performance, the average expected estimation error, the associated sufficient conditions are also derived based on a different analysis approach with the existing work. And a feasible method is presented for both indexes to seek the optimal constant attack power level when the system fails to meet the proposed sufficient conditions. Then when the real-time ACK information can be acquired, the attacker desires a time-varying power attack strategy, based on which a Markov decision process (MDP) based algorithm is designed to solve the optimal dynamic attack energy management problem. We further study the optimal tradeoff between attack energy and system degradation. Specifically, by moving the energy constraint into the objective function to maximize the system index and minimize the energy consumption simultaneously, the other MDP based algorithm is proposed to find the optimal dynamic attack power policy which is further shown to have a monotone structure. The theoretical results are illustrated by simulations.
\end{abstract}
\section{Introduction}
Cyber-Physical Systems (CPS) tightly conjoin cyber
elements and physical resources for sensing, control, communication, and computation. CPS, regarded as the next generation engineered systems, are common in a large scope of infrastructures, such as transportation systems, autonomous vehicles, smart buildings and mine monitoring, in many of which safety is crucial \cite{Johansson2014,Poovendran2012}. Nevertheless, in light of the nature of the high openness, CPS are vulnerable to the malicious threats from the external. This has triggered a great deal of attention to the issues of \emph{cyber-security} \cite{Fawzi2014,Sundaram2011,Pajic2014,Mo2015, Vamvoudakis2014,Brown2017,Amini2018,Jiang2018}.

Researchers investigated \emph{cyber-security} under some specific attack patterns. Different attack patterns include deception attacks \cite{Amin2013a,Amin2013b}, replay attacks \cite{Mo2009,Zhu2014}, false data injection attacks \cite{Mo2014}, and Denial-of-Service (DoS) attacks \cite{Befekadu2015,Peris2015,Foroush2012}.
In the current paper, we focus on DoS attacks. The adversary launches DoS attacks to jam a wireless channel through which useful information is transmitted. In wireless communications, several reasons, including channel fading, scattering, signal degradation, etc, lead to the random dropout of data packets. Due to the interference from the attacker, the packet drops with a larger probability. Specifically, higher jamming power leads to a smaller signal-to-interference-plus-noise ratio (SINR), which implies a higher probability of the packet dropout \cite{Poisel2011}.

It has been pointed out that a DoS attacker may be subject to a limited energy/power budget, i.e., the attacker cannot block the communication network ceaselessly \cite{Li2015b,Zhang2015,Zhang2016}. Zhang et al. \cite{Zhang2015} investigated how an attacker with limited ability schedules the DoS attacks to maximize the estimation error at the remote estimator. They also proposed the optimal DoS attack policies to maximize the Linear Quadratic Gaussian control cost function when the attacker cannot deteriorate the channel quality all the time \cite{Zhang2016}. Li et al. proposed a zero-sum game in which the sensor and the DoS attacker, both with energy constraint, find their optimal mixed strategies to maximize their payoff functions, and demonstrated that the optimal strategies for these two players constitute a mixed strategy Nash Equilibrium in\cite{Li2015b}.
All these works assumed that the attacker can only obstruct the channel $m$ times over a finite time horizon $T$ ($m<T$), and focused on when the attacker should jam the channel. However, they did not consider which power level the DoS attacker should employ.

When taking into account the power level of the DoS attack, the issue of finding the optimal power level that most severely degrades the system performance follows.
The recent work in \cite{Zhang2018} considered DoS attack energy management under the constraints of restricted attack power over a finite time horizon. The local estimate of a sensor is sent to a remote estimator through a wireless link under DoS attacks. It is assumed that the same attack energy is used each time. They proved that, under the same attack times, the higher attack power, the larger the trace of expected terminal estimation error covariance. Due to the fact that higher attack energy leads to the smaller attack times, the attacker has to decide how much energy to employ when launching a DoS attack. For a special system case (the system matrix is normal), the authors provided some sufficient conditions under which the analytical solution to the optimal static attack energy management problem is obtained, but failed to derive the closed-form solution for a general system case. Zhang et al. \cite{Zhang2017} also studied the other system performance index, the average expected estimation error, in the setup of \cite{Zhang2018} and obtained the corresponding sufficient conditions. In both \cite{Zhang2018} and \cite{Zhang2017}, the model of packet dropout probability is based on the signal-to-interference-plus-noise ratio at the receiver, in which different attack power levels correspond to different probabilities of packet dropout, including the level at which the attack power is zero, i.e., there is no attack. However, like most existing works that discard the intermittent packet dropout in the absence of attack \cite{Befekadu2015,Peris2015,Zhang2016,Zhang2015,
Foroush2012}, Ref. \cite{Zhang2018} and \cite{Zhang2017} also assumed that data packets will reach the remote estimator if DoS attack does not appear, which is against the SINR-based model and strongly restrict the application in the real scenario. In contrast, our previous work \cite{Qin2018} considered the scenario of remote state estimation under DoS attacks in which the random losses may occur even if there is no attack. Therein, the problem of when the attacker should jam the channel is solved, but the effect of the attack power on remote estimation performance is neglected.

In the current work, to capture the packet dropout in wireless channels and have an insight of how different attack power levels would impact on the system performance, we investigate the optimal DoS attack energy management problem over an SINR-based network which embeds the works in \cite{Zhang2015,Zhang2018,Zhang2017,Qin2018} as special cases.
The main contributions of our work consist of the following.
\begin{enumerate}
  \item For a general system case (not confined to the system with a normal system matrix which the result in \cite{Zhang2018} requires), we propose new sufficient conditions under which the closed-form solution to the optimal static attack energy management problem for the expected terminal estimation error is acquired. We further prove that the obtained sufficient conditions are more relaxed than the ones in \cite{Zhang2018}, i.e., the new conditions hold if the ones in \cite{Zhang2018} are met.
  \item Based on the network model of \cite{Zhang2017}, the proposed problem can be easily tackled. However, their proof technique does not apply to the analysis in our setup due to the introduction of the SINR-based network. In this paper, by inducing a virtual random variable from the definition of the average expected estimation error, we are capable of using an usual stochastic order inequality argument to proceed to the analysis. Then the associated sufficient conditions are derived for the average expected estimation error. When the conditions are not satisfied, we show that the algorithm proposed in \cite{Zhang2018} and \cite{Zhang2017} is feasible for both indexes in our setup.
  \item Optimal dynamic attack energy management problems for both indexes of system performance are solved by a Markov decision process (MDP) based algorithm. We further consider the optimal tradeoff problem between attack energy and system degradation, the solution of which can be acquired by the other MDP-based algorithm. And the corresponding solution is found to have a monotone structure, which significantly reduce the computational complexity of the proposed algorithm.
\end{enumerate}

The remainder of the paper is organized as follows: Section II formulates the problem of interest. In section III, we focus on the expected terminal estimation error. We present some sufficient conditions under which the optimal attack power level can be obtained explicitly. Then we prove that our proposed conditions are more relaxed than the existing work. Methods are presented to solve the optimal static attack energy management problem when the sufficient conditions fail to hold. Section IV derives the results for the average expected estimation error. The corresponding sufficient conditions and closed-form attack power level are derived. Corresponding methods are also provided when the sufficient conditions are not met. In section V, we formulate the optimal dynamic attack energy management problem for both indexes and design a MDP-based algorithm to solve it. Section VI considers the tradeoff between attack energy and system degradation. The optimal solution which offers best tradeoff can be obtained by the other MDP-based algorithm and is shown to have a monotone structure. Numerical simulations are provided in section VII to demonstrate the theoretical results. Finally, some concluding remarks appear in section VIII.

\emph{Notations:} $S_+^n$ is the set of $n$ by $n$ positive semi-definite matrices. $X_1\leq X_2$ means $(X_2-X_1)\in S_+^n$, and $X_1>0$ denotes that $X_1$ is positive definite. Denote by $\mathbb{Z}^+$ the set of positive integers. $Pr(\mathcal{A})$ stands for the probability of an event $\mathcal{A}$ and $Pr(\mathcal{A}|\mathcal{B})$ for the conditional probability given an event $\mathcal{B}$. The mean of random variable $X$ is denoted as $\mathbb{E}[X]$, and $\mathbb{E}[X|\mathcal{A}]$ is the conditional expectation of $X$ given an event $\mathcal{A}$. $\textrm{Tr}(\cdot)$ denotes the trace of matrix. The superscript ${'}$ stands for transposition. For function $h$, $h^{k}(X)\triangleq h(h^{k-1}(X))$, with $h^0(X)\triangleq X$.

\section{Problem Setup}\label{sec:system}
The system in Fig. 1 is considered. A linear time-invariant (LTI) process is run by the plant, and the sensor takes the measurement of the state in the plant, as follows.
\begin{align*}
x_{k+1}&=Ax_{k}+w_{k},\\
y_{k}&=Cx_{k}+v_{k},
\end{align*}
where $k\in \mathbb{Z}^+$, $x_{k}, w_{k}\in \mathbb{R}^{n}$ and $y_{k}, v_{k}\in \mathbb{R}^{m}$ are the process state, the process noise that is zero-mean Gaussian noise with covariance $Q\geqslant0$, the measurement taken by the sensor, and the measurement noise that is zero-mean Gaussian noise with covariance $R>0$, respectively, at time $k$. Furthermore, $w_{k}$ and $v_{k}$ are uncorrelated. Assume that the pair $(A,C)$ is detectable and $(A,\sqrt{Q})$ is controllable.

At time $k$, after obtaining the measurement data $y_k$, the sensor with sufficient computation ability runs a Kalman filter which generates the minimum mean squared error (MMSE) estimate $\hat{x}_k^s=\mathbb{E}[x_k|y_1,\ldots,y_k]$, with the corresponding error covariance $P_k^s=\mathbb{E}[(x_k-\hat x_k^s)(x_k-\hat x_k^s)^{'}|y_1,\ldots,y_k]$.
Then the data packet, $\hat{x}_k^s$, is transmitted from the sensor with the power $\delta^s$ to a remote estimator over a wireless network. Consider the state estimation at the remote
estimator within a finite time horizon $T$, with the wireless channel under DoS attacks $\delta\buildrel{\Delta}\over{=}\{\delta_{1},
\delta_{2},\ldots,\delta_{T}\}$, where $\delta_{k}$ is the attack power at time $k$. Let $\lambda_k=1$ if $\delta_{k}>0$, and $\lambda_k=0$ if $\delta_{k}=0$. Assume the same power is employed when jamming the channel, i.e., $\delta_{k}=\delta^a$ if $\lambda_k=1$. Then from \cite{Zhang2015}, the following equality describes the estimation procedure of the state estimate $\hat{x}_{k}$ at the remote estimator:
\begin{align*}
\hat{x}_{k} = \theta_k(\delta)\hat{x}_{k}^s+ (1-\theta_k(\delta))A\hat{x}_{k-1},\,\,\, k=1, 2, \ldots, T,
\end{align*}
where $\theta_k =1$ if the packet arrives at the estimator at time $k$ and $\theta_k =0$, otherwise. The DoS attacker has a limited energy resource, which is formulated as $\sum_{k=1}^{T}\delta_{k}\leq \Delta$.When $\delta$ is given, $\theta_k$'s are assumed to be i.i.d. Bernoulli random variables, and the corresponding probability distribution follows
\begin{align}\label{pk}
\tilde{p}_k(\delta)=Pr(\theta_k =1)= 1-\beta_k,
\end{align}
where $\beta_k$ is the packet dropout probability under the DoS attack with the power $\delta_k$. The packet dropout probability at time $k$ is decided by the SINR of the remote estimator at time $k$ which is defined by \cite{Poisel2011}
\begin{align}\label{sinr}
\rho_k=\frac{\delta^sG^s}{\delta_kG^a+\sigma^2},
\end{align}
where $G^s$ and $G^a$ are the channel gains for the sensor and the attacker respectively and $\sigma^2$ is the noise power.

Assume that the packet is composed with $L$ bits, and the bit error rate for each bit is identical. The packet dropout takes place if there exists one bit that is received mistakenly. Then from \cite{Poisel2011}, there holds
\begin{align}\label{pksinr}
\tilde{p}_k=[1-\mathcal{Q}(\sqrt{2\rho_k})]^L,
\end{align}
where $\mathcal{Q}(x)=1/\sqrt{2\pi}\int_x^{+\infty}
e^{-t^2/2}dt$. And $\beta_k=1-\tilde{p}_k$.

In the Kalman filter, $P_k^s$ will converge exponentially to the steady-state value, $\overline{P}$. Similar to \cite{Li2015b}, assume the Kalman filter is in steady-state, which implies $P_k^s=\overline{P}, k\geq 0$.
Then according to \cite{Zhang2015}, the error covariance $P_{k}$ of $\hat{x}_{k}$ follows:
\begin{align}\label{Pk}
P_{k} = \theta_k\overline{P}+ (1-\theta_k)h(P_{k-1}),\,\,\, k=1, 2, \ldots, T,
\end{align}
where $h(X)\triangleq{AXA^{'}+Q}$. And similar to \cite{Qin2018}, it is assumed throughout the paper that the data packet which contains the information of $\hat{x}_{0}^{s}$ successfully arrives at the remote estimator at time $k=0$, i.e., $P_0=\overline{P}$.

Two types of indexes are adopted to measure the performance of remote estimation in existing works \cite{Li2015b,Zhang2015,Zhang2018,Savage2009,Shi2012}. They are respectively the expected terminal estimation error covariance called \emph{Terminal Error} $J^T=\mathbb{E}[P_T(\delta)]$, and the average expected estimation error covariance called \emph{Average Error} $J^A=\frac1T\sum_{k=1}^T\mathbb{E}[P_k(\delta)].$ The attacker subject to the limited energy budget expects to deteriorate the remote estimation performance as much as possible:
\begin{figure}
  \centering
  \includegraphics[width=0.47\textwidth]{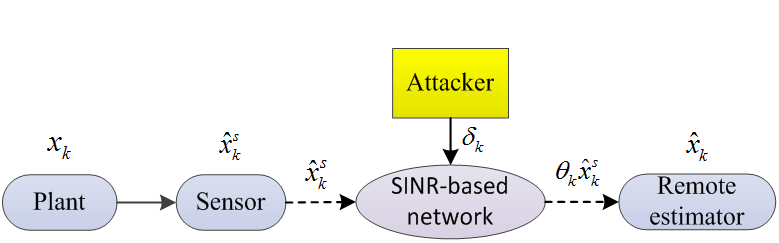}\\
  \caption{System architecture.}
\end{figure}

\emph{Problem 1:}
\begin{eqnarray*}
& & \max_{\delta\in\Theta} \quad \textrm{Tr}[J^e(\delta)] \\
& & s.t. \quad\sum_{k=1}^{T}\delta_{k}\leq \Delta,\\
& & \qquad \,\,\,\,\underline{\delta}\leq \delta^a \leq \overline{\delta},
\end{eqnarray*}
where, $J^e=J^T$ or $J^e=J^A$, $\Theta=\{0,\delta^a\}^T$ is the set of all possible attack power allocations, and $\delta=\{\delta_{1},
\delta_{2},\ldots,\delta_{T}\}$, $\underline{\delta}$ and $\overline{\delta}$ are the
lower bound and upper bound of attack power, respectively.

\section{Optimal Static Attack Energy Management For Terminal Error}
We first work on the terminal error. In this section, the optimal attack energy management for maximizing the trace of the terminal error is presented. The analysis of \emph{Problem 1} for the terminal error in detail is provided.
\subsection{Optimal Attack Schedule For Terminal Error}
When the constant attack power $\delta^a$ is given, the attack times can be computed by $n=\lfloor\Delta/\delta^a\rfloor$. We denote the probability of packet dropout by $\beta$ when $\delta_k=\delta^a$, and $\alpha$ when $\delta_k=0$. Then, for a given $\delta^a$, we need first solve the problem of when to jam the channel:

\emph{Problem 2:}
\begin{eqnarray*}
& & \max_{\lambda\in\Lambda} \quad \textrm{Tr}[J^T(\lambda)] \\
& & s.t. \quad\sum_{k=1}^{T}\lambda_{k}= n,
\end{eqnarray*}
where, $\Lambda=\{0,1\}^T$ is the set of all possible attack schedules, and $\lambda=\{\lambda_{1},
\lambda_{2},\ldots,\lambda_{T}\}$.

From \cite{Qin2018},  the optimal attack schedule for \emph{Problem 2} is given by the following lemma.

\emph{Lemma 1 (\cite{Qin2018}):}
The optimal solution to \emph{Problem 2} is
$$\lambda^*=(0,0,\cdots,0,\underbrace{1,1,\cdots,1}
_{n\,\, {\rm times}}),$$
and the trace of corresponding expected terminal estimation
error is
\begin{align}\label{Jmax}
\textrm{Tr}(J_{max}^T)= &\,\textrm{Tr}\Big[\sum_{i=0}^{n-1}(1-\beta)\beta^{i}h^{i}
(\overline{P})+\alpha^{T-n}\beta^{n}h^{T}
  (\overline{P})\nonumber\\
  &+\sum_{i=n}^{T-1}(1-\alpha)
  \alpha^{i-n}\beta^{n}   h^{i}(\overline{P})\Big]
\end{align}
\subsection{Sufficient Conditions}
It is difficult to give a closed-form solution to \emph{Problem 1} for the terminal error if no restriction is imposed on the system parameters. This subsection presents some sufficient conditions under which the closed-form solution can be derived.

When the probability of packet dropout $x$ is given, we can obtain the corresponding attack power $\delta^a(x)$ based on \eqref{pk}--\eqref{pksinr}. Denote $\mathcal{S}_n=\{x|\lfloor\Delta/\delta^a(x)
\rfloor=n\}$ as the set of the
packet dropout probability with the capability of launching $n$ times attacks for the adversary. Before proceeding further, the following two lemmas are needed.

\emph{Lemma 2} (\cite{Zhang2015} ): For function $h$ defined in Section II, the following holds
$$\overline{P}\leq h(\overline{P})\leq h^{2}(\overline{P})\leq\cdots\leq h^{k}(\overline{P})\leq\cdots,\forall k\in {\mathbb{Z}}^{+}.$$

\emph{Lemma 3:} If $\beta_1, \beta_2\in \mathcal{S}_n$, and
$\beta_1>\beta_2$, then the attack strategy in \emph{Lemma 1} is optimal for $\beta_1$ and $\beta_2$. And there holds
\begin{align*}
\textrm{Tr}(J_{max}^T)(\beta_1)>\textrm{Tr}(J_{max}^T)(\beta_2).
\end{align*}
\begin{IEEEproof}
According to \eqref{Jmax}, it follows that
\begin{align}\label{Jmaxn}
J_{max}^T=\overline{P}+\sum_{i=1}^{n}\beta^iH^i+\sum_{i=n+1}^{T}\beta^n\alpha^{i-n}H^i,
\end{align}
where $H^i=h^{i}(\overline{P})-h^{i-1}(\overline{P})$.

From \eqref{Jmaxn} and \emph{Lemma 2}, it is easy to see that
\begin{align*}
\textrm{Tr}(J_{max}^T)(\beta_1)>\textrm{Tr}(J_{max}^T)(\beta_2),
\end{align*}
which completes the proof.
\end{IEEEproof}

Then the following proposition shows the effect of attack times on the system performance measured by the terminal error. Specifically, more attack times lead to larger trace of the terminal error if some conditions are satisfied.

\emph{Proposition 1:} Let $\overline{n}=\lfloor\Delta/\underline{\delta}
\rfloor$ and $\underline{n}=\lfloor\Delta/\overline{\delta}\rfloor$.
Suppose $\theta\in\mathcal{S}_n$ and $\beta\in\mathcal{S}_{n+1}$, where $\underline{n}\leq n\leq \overline{n}-1$.
If the following conditions are met:
\begin{enumerate}
  \item $\forall t_1, t_2\in {\mathbb{Z}}^{+}$, and $t_1\leq t_2$, there holds
$\textrm{Tr}(H^{t_1})\leq \textrm{Tr}(H^{t_2}).$
  \item the probabilities satisfy
  \begin{align*}
\sum_{i=1}^{\overline{n}-1}(\overline{\beta}^i-\underline{\beta}
^i)+(\alpha\overline{\beta}^{\overline{n}-1}-\underline{\beta}
^{\overline{n}})\sum_{i=0}^{T-\overline{n}}\alpha^{i}\leq 0,
\end{align*}
where $\underline{\beta}$ and $\overline{\beta}$ are the lower bound
and upper bound of packet
dropout probability, which are corresponding to the upper bound
and lower bound of the attack power, respectively,
\end{enumerate}
then we have
\begin{align*}
\textrm{Tr}(J_{max}^T)(\theta)\leq \textrm{Tr}(J_{max}^T)(\beta).
\end{align*}
\begin{IEEEproof}
See the Appendix.
\end{IEEEproof}
\subsection{Closed-form Solution}
Now the main conclusion is provided as follows.

\emph{Theorem 1:} Let $\overline{n}=\lfloor\Delta/\underline{\delta}\rfloor$ and $\underline{n}=\lfloor\Delta/\overline{\delta}\rfloor$. If conditions 1) and 2) in \emph{Proposition 1} are satisfied, then the optimal attack power level for the terminal error is
\begin{align}\label{the}
\delta^{a}_*= \left\{ \begin{array}{ll}
\textrm{arg max $\{\delta^{a}|\lfloor\Delta/\delta^{a}\rfloor=\overline{n}\}$} ,& \textrm{for $\underline{n}<\overline{n}$}, \\
\overline{\delta} ,& \textrm{for $\underline{n}=\overline{n}$},
\end{array} \right.
\end{align}
the solution to \emph{Problem 1} for the terminal error is
$$\delta{\ast}=(0,0,\cdots,0,\underbrace{\delta^{a}_*,
\delta^{a}_*,\cdots,\delta^{a}_*}_{\overline{n}\,\, {\rm times}}),$$
and the trace of the corresponding expected terminal error covariance is
\begin{align*}
\textrm{Tr}(J_{max}^T)= &\,Tr\Big[\sum_{i=0}^{\overline{n}-1}(1-\beta_*)\beta^{i}
_*h^{i}(\overline{P})+\alpha^{T-\overline{n}}\beta
^{\overline{n}}_*h^{T}(\overline{P})\nonumber\\
  &+\sum_{i=\overline{n}}^{T-1}(1-\alpha)
  \alpha^{i-\overline{n}}\beta^{\overline{n}}_*             h^{i}(\overline{P})\Big],
\end{align*}
where $\beta_*=1-\tilde{p}_k(\delta^{a}_*)$ is the packet dropout probability corresponding to the optimal DoS attack power.
\begin{IEEEproof}
To maximize the terminal error, due to \emph{Lemma 1} and \emph{Proposition 1}, the attacker will block the channel $\overline{n}$ times. From \emph{Lemma 3}, the attacker will employ the larger power level. Hence, the conclusion in \emph{Theorem 1} holds.
\end{IEEEproof}

From \emph{Theorem 1}, the attacker should adopt the largest power level among the power level set in which any power level leads to most attack times, and then consecutively jam the channel in the end of the considered time horizon to maximize the terminal error when the proposed sufficient conditions are satisfied.

For any given plant, satisfaction of the condition
1) in \emph{Proposition 1} can be checked. However, it is not easy to check for large dimension of the system matrix, which motivates the sufficient condition in the following corollary.

\emph{Corollary 1:} For function $h$ defined in Section II, the condition 1) of \emph{Proposition 1} holds if $\lambda_{min}(A'A)\geq 1$, where $\lambda_{min}(A'A)$ is the minimum eigenvalue of $A'A$.

\begin{IEEEproof}
It suffices to prove the case that $t_2=t_1+1$.

Since
\begin{align}\label{recu}
&h^{t_1+1}(\overline{P})-h^{t_1}(\overline{P})=
Ah^{t_1}(\overline{P})A'-Ah^{t_1-1}(\overline{P})A'\nonumber\\
=&A[h^{t_1}(\overline{P})-h^{t_1-1}(\overline{P})]A',
\end{align}
based on the facts that $\textrm{Tr}(ABC)=\textrm{Tr}(CAB)$ and $\textrm{Tr}(A+B)=\textrm{Tr}(A)+\textrm{Tr}(B)$, there holds
\begin{align}\label{dif}
&\textrm{Tr}\{h^{t_1+1}(\overline{P})-h^{t_1}(\overline{P})\}-
\textrm{Tr}\{h^{t_1}(\overline{P})-h^{t_1-1}(\overline{P})\}\nonumber\\
=&\textrm{Tr}\{A'A[h^{t_1}(\overline{P})-h^{t_1-1}(\overline{P})]\}-
\textrm{Tr}\{h^{t_1}(\overline{P})-h^{t_1-1}(\overline{P})\}\nonumber\\
=&\textrm{Tr}\{(A'A-I)[h^{t_1}(\overline{P})-h^{t_1-1}(\overline{P})]\}.
\end{align}
Since $\lambda_{min}(A'A)\geq 1$, we have $A'A-I \in S_+^n$, which causes, from \emph{Lemma 1}, that
\begin{align*}
&\textrm{Tr}\{h^{t_1+1}(\overline{P})-h^{t_1}(\overline{P})\}-
\textrm{Tr}\{h^{t_1}(\overline{P})-h^{t_1-1}(\overline{P})\}\geq 0.
\end{align*}
The proof is completed.
\end{IEEEproof}

Note that \emph{Corollary 1} provides a sufficient condition under which the condition 1) of \emph{Proposition 1} holds, i.e., the condition 1) of \emph{Proposition 1} may hold even if $\lambda_{min}(A'A)<1$. This is illustrated by the following example.

\emph{Example 1:} $A=\begin{bmatrix} 1.2 & 0.1 \\ 0 & 1 \end{bmatrix}$, $C=\begin{bmatrix} 1 & 0 \\ 0 & 1 \end{bmatrix}$, $Q=\begin{bmatrix} 1 & 0 \\ 0 & 2 \end{bmatrix}$, $R=0.5C$. The eigenvalues of $A'A$ are $0.9788$ and $1.4712$. Hence we have $\lambda_{min}(A'A)<1$. We can run a Kalman filter to obtain the steady-state error covariance $\overline{P}$, and compute the difference
\begin{align}\label{dif2}
h(\overline{P})-\overline{P}=\begin{bmatrix} 1.1709  &  0.0418\\
    0.0418  &  2.0000 \end{bmatrix}.
\end{align}
Similar to \eqref{recu}, there holds
\begin{align}\label{recu2}
h^{t_1}(\overline{P})-h^{t_1-1}(\overline{P})
=A^{t_1-1}[h(\overline{P})-\overline{P}](A^{t_1-1})'.
\end{align}
It is easy to verify that each element of the matrix $A'A-I$ is greater than $0$, which leads to, from \eqref{dif}--\eqref{recu2},
\begin{align*}
&\textrm{Tr}\{h^{t_1+1}(\overline{P})-h^{t_1}(\overline{P})\}-
\textrm{Tr}\{h^{t_1}(\overline{P})-h^{t_1-1}(\overline{P})\}\nonumber\\
=&\,\textrm{Tr}\{(A^{t_1-1})'(A'A-I)A^{t_1-1}[h(\overline{P})-\overline{P}]\}
\geq 0.
\end{align*}
\subsection{Comparison with Existing Results}
In this subsection, \emph{Theorem 1} is compared with the existing results in \cite{Zhang2018}.

\emph{Definition 1 (Normal matrix \cite{Horn2012}):} A square matrix $A$ is normal if $A^*A=AA^*$, where $A^*$ is the conjugate transpose of $A$.

\emph{Lemma 4 (\cite{Zhang2018}):} When $\alpha=0$, the optimal attack power level for the terminal error is \eqref{the} if the following conditions are satisfied:
\begin{enumerate}
  \item matrix $A$ is normal,
  \item all the eigenvalues of the matrix $A'A$, $\lambda(A'A)$, satisfy\footnote{There is a typo in \cite{Zhang2018}.}
      \begin{align}\label{cond}
1<\lambda(A'A)<\frac1{\overline{\beta}}[1-(\frac{\overline{\beta}
-\underline{\beta}}{\underline{\beta}^{\overline{n}+1}})^{1/2}]
.
\end{align}
\end{enumerate}

Next we show that the conditions in \emph{Proposition 1} are more relaxed than the ones in \emph{Lemma 4}.

We have $\lambda(A'A)>1$, if \eqref{cond} holds. Then from \emph{Corollary 1}, it is easy to see that, the condition 1) of \emph{Proposition 1} is true if the condition 2) of \emph{Lemma 4} holds.

When $\alpha=0$, the condition 2) in \emph{Proposition 1} turns into
\begin{align}\label{condus}
\sum_{i=1}^{\overline{n}-1}(\overline{\beta}^i-\underline{\beta}
^i)-\underline{\beta}^{\overline{n}}\leq 0.
\end{align}

Since
\begin{align*}
\overline{\beta}^i-\underline{\beta}^i=(\overline{\beta}
-\underline{\beta})\sum_{k=0}^{i-1}\overline{\beta}^{i-k-1}
\underline{\beta}^k
\leq(\overline{\beta}-\underline{\beta})i\overline{\beta}^{i-1},
\end{align*}
there holds
  \begin{align}\label{ine}
&\sum_{i=1}^{\overline{n}-1}(\overline{\beta}^i-\underline{\beta}
^i)-\underline{\beta}^{\overline{n}}\leq(\overline{\beta}
-\underline{\beta})\sum_{i=1}^{\overline{n}-1}i\overline{\beta}
^{i-1}\nonumber\\
=&(\overline{\beta}-\underline{\beta})
[\frac{1-\overline{\beta}^{\overline{n}-1}}{(1-\overline{\beta})
^2}-\frac{(\overline{n}-1)\overline{\beta}^{\overline{n}-1}}
{1-\overline{\beta}}]-\underline{\beta}^{\overline{n}},
\end{align}

If \eqref{cond} is met, it follows that
\begin{align*}
1<\frac1{\overline{\beta}}[1-(\frac{\overline{\beta}
-\underline{\beta}}{\underline{\beta}^{\overline{n}+1}})^{1/2}],
\end{align*}
which amounts to
\begin{align*}
\frac{1}{(1-\overline{\beta})^2}<\frac{\underline{\beta}
^{\overline{n}+1}}{\overline{\beta}-\underline{\beta}}.
\end{align*}

Then in light of the above inequality, the last term in \eqref{ine} is smaller than
\begin{align*}
\underline{\beta}^{\overline{n}+1}-\underline{\beta}^{\overline{n}
}-\underline{\beta}^{\overline{n}+1}\overline{\beta}^{\overline{n}
-1}-\frac{(\overline{n}-1)\overline{\beta}^{\overline{n}-1}
(\overline{\beta}-\underline{\beta})}{1-\overline{\beta}}\leq 0,
\end{align*}
which leads to, due to \eqref{ine}, the inequality \eqref{condus}.

Therefore, the condition 2) in \emph{Proposition 1} is satisfied if the condition 2) in \emph{Lemma 4} holds.

\emph{Remark 1:} To summarize, the conditions in \emph{Proposition 1} hold if the conditions in \emph{Lemma 4} are satisfied, which means that the conditions in \emph{Proposition 1} are more relaxed. Besides, from the derivation above, just two inequalities, $\lambda(A'A)>1$ and $\frac1{\overline{\beta}}[1-(\frac{\overline{\beta}
-\underline{\beta}}{\underline{\beta}^{\overline{n}+1}})^{1/2}]>1$, are employed. Hence, compared with \cite{Zhang2018}, to obtain the closed-form solution, it is no longer required that the system matrix $A$ is normal, and no upper bound is imposed on $\lambda(A'A)$, which largely improves the applicability.

\subsection{Exhaustion Search Method}
When the conditions in \emph{Proposition 1} are not met,  \eqref{the} may not be the optimal one. But from \emph{Lemma 3}, we can transform \emph{Problem 1} for the terminal error into an equivalent problem:

\emph{Problem 3:}
\begin{eqnarray*}
& & \max \quad\textrm{Tr}[J_{max}^T(\delta)] \\
& & s.t. \quad (\delta^a,M)\in \Omega,
\end{eqnarray*}
where, $\delta=(0,0,\cdots,0,\underbrace{\delta^{a},
\delta^{a},\cdots,\delta^{a}}_{M\,\, {\rm times}})$, $\Omega=\Big\{(f(n),n), n=\underline{n},\underline{n}+1,\cdots,\overline{n}\Big\}$, and $f(n)=\textrm{arg max}\{x|\lfloor\Delta/x\rfloor=n,\underline{\delta}\leq x\leq \overline{\delta}\}$.

It is easy to see that the solution space of \emph{Problem 3} is discrete. For a given $(\delta^a,M)\in \Omega$, we can compute the corresponding $\alpha$ and $\beta$ based on \eqref{pk}--\eqref{pksinr}. Then $\textrm{Tr}[J_{max}^T(\delta)]$ can be acquired from \eqref{Jmax}. Finally, an exhaustion search method can be adopted to solve \emph{Problem 3}. Note that, from the constraint set $\Omega$, no more than $T$ steps are required to obtain the solution, which implies that this method is feasible.

\section{Optimal Static Attack Energy Management For Average Error}
In this section, we focus on the other important system index, the average error. The optimal attack energy management for maximizing the trace of the average error is presented. We also provide the detailed analysis of \emph{Problem 1} for the average error.
\subsection{Optimal Attack Schedule For Average Error}
The scenario for the average error is far more complicated than the one for the terminal error. To facilitate the subsequent analysis, we first introduce the representation of the attack schedule. An attack schedule, in which $n$ attacks are launched over a finite horizon $T$, can be denoted by
\begin{align*}
(\gamma^{d},\lambda^{k_1},\gamma^{d_1},\lambda^{k_2},\ldots,
\lambda^{k_{s}},\gamma^{d_s}),
\end{align*}
where $s\geq 1$, $\lambda^{k_i}$ is the $i$th consecutive jamming sequence with the length $k_i\geq1$,  $i=1,\ldots,s$, $\gamma^{d}$ and $\gamma^{d_j}$, respectively, denote the first and the $(j+1)$th consecutive sequence during which no attack is launched with the length $d_j\geq1$, $j=1,\ldots,s-1$, $d\geq0$ and $d_s\geq0$, i.e.,
$$(\underbrace{0, \ldots, 0}_{d \,{\rm times}}, \underbrace{1,\ldots,1}_{k_{1}\,{\rm times}},0, \ldots, 0, \underbrace{1,\ldots,1}_{k_{s}\,{\rm times}},\underbrace{0, \ldots,0}_{d_{s}\,{\rm times}}),$$
where $\sum_{i=1}^{s}k_i= n$, and $d+\sum_{j=1}^{s}d_j= T-n$.

Similar to Section III, we first solve the following problem:

\emph{Problem 4:}
\begin{eqnarray*}
& & \max_{\lambda\in\Lambda} \quad \textrm{Tr}[J^A(\lambda)] \\
& & s.t. \quad\sum_{k=1}^{T}\lambda_{k}= n,
\end{eqnarray*}
where, $\Lambda=\{0,1\}^T$ is the set of all possible attack schedules, and $\lambda=\{\lambda_{1},
\lambda_{2},\ldots,\lambda_{T}\}$.

From \cite{Qin2018},  the optimal attack schedule for \emph{Problem 4} is given by the following lemma.

\emph{Lemma 5 (\cite{Qin2018}):}
The solution to \emph{Problem 4} is $\lambda_{\ast}=(\gamma^{m},\lambda^{n},\gamma^{s})$, i.e.,
$$\lambda_{\ast}=(\underbrace{0,0,\ldots,0}_{m\, {\rm times}},\underbrace{1,1,\ldots,1}_{n\, {\rm times}},\underbrace{0,0,\ldots,0}_{s\, {\rm times}}),$$
where $m+s=T-n$, and $\vert{m-s}\vert\leq1$, i.e., $m=s$ or $\vert{m-s}\vert=1$. And $J^A[(\gamma^{m},\lambda^{n},\gamma^{s})]
=J^A[(\gamma^{s},\lambda^{n},\gamma^{m})]$. Denote by $p_{i,k}$ the probability that $P_k=h^{i}(\overline{P})$, $i=0,1,\ldots,T$. Then the corresponding average expected estimation error can be calculated based on the expression $$J_{max}^A=\frac1T\sum_{k=1}^T\sum_{i=0}^{T}p_{i,k}(\lambda_{\ast})h^{i}(\overline{P}).$$

To obtain $p_{i,k}$, from \eqref{Pk} and the assumption that $\hat{x}_{0}^{s}$ arrives, it follows that  $p_{i,k}=0$ when $k<i$. When $k\geq i$, $P_k=h^{i}(\overline{P})$ if and only if $\hat{x}_{k-i}^{s}$ arrives, and meanwhile $\hat{x}_{k-i+1}^{s}, \hat{x}_{k-i+2}^{s}, \ldots, \hat{x}_{k}^{s}$ drop. Note that $\theta_k$'s are independent.
Hence, $p_{i,k}=\tilde{p}_{k-i}
\tilde{q}_{k-i+1}\tilde{q}_{k-i+2}\cdots\tilde{q}_k$ when $k\geq i$, where $\tilde{q}_k=1-\tilde{p}_k$ is the corresponding dropout probability of packets. And there holds $\tilde{p}_0= 1$ due to the assumption that $\hat{x}_{0}^{s}$ arrives.

In \cite{Zhang2017}, $J_A(\lambda)$ has a simple and tractable expression because of the assumption that the packet dropout probability without attack is $\alpha=0$. In our setup, however, the corresponding analysis is more difficult since $\alpha >0$. Hence, we need a different analysis approach from the one in \cite{Zhang2017} to tackle \emph{Problem 1} for the average error, as will be presented in the next subsection.
\subsection{Sufficient Conditions}
For ease of illustration, we rewrite the optimal attack schedule in \emph{Lemma 5} as $\phi^1=(\gamma^{m},\lambda^{n},\gamma^{s})$, and let $\phi^2=(\gamma^{m},\lambda^{n+1},\gamma^{s-1})$. To guarantee that $\phi^2$ is the optimal attack schedule for the average error under $n+1$ attacks, let $m=s$ or $m=s-1$.

We aim to compare the effects of two kinds of attack schedules $\phi^1$ and $\phi^2$ on the system performance, which help us have an insight of the effect of attack times. To do this, we have, from \emph{Lemma 5},
\begin{align}\label{JAD}
J_{max}^A(\phi^1)-J^A_{max}(\phi^2)=\frac1T\sum_{i=0}^{T}h^{i}(\overline{P})F_i,
\end{align}
where $F_i=\sum_{k=1}^T[p_{i,k}(\phi^1)-p_{i,k}(\phi^2)]$.

Observing the above equation, two random variables can be induced from \eqref{JAD}. First, note that $\{\frac1T\sum_{k=1}^Tp_{i,k}, i=0,1,\ldots,T\}$ is a probability distribution associated with some random variable since $0\leq \frac1T\sum_{k=1}^Tp_{i,k}\leq 1, i=0,1,\ldots,T$ and $\sum_{i=0}^{T}\frac1T\sum_{k=1}^Tp_{i,k}=1$. Then we can obtain two induced random variables $X$ with probability distribution $Pr(X=\textrm{Tr}[h^{i}(\overline{P})])=\frac1T\sum_{k=1}^Tp_{i,k}(\phi^1), i=0,1,\ldots,T$ and $Y$ with probability distribution $Pr(Y=\textrm{Tr}[h^{i}(\overline{P})])=\frac1T\sum_{k=1}^Tp_{i,k}(\phi^2), i=0,1,\ldots,T$. And we readily have $\mathbb{E}[X]=\textrm{Tr}(J_{max}^A)(\phi^1)$ and $\mathbb{E}[Y]=\textrm{Tr}(J_{max}^A)(\phi^2)$.

Before proceeding further, we present the following definition and lemma.

\emph{Definition 2 (\cite{Shaked2007} ):} Let $X^1$ and $X^2$ be two random variables. $X^1$ is said to be smaller than $X^2$ in the usual stochastic order (denoted by $X^1\leq_{\textrm{st}} X^2$) if $Pr(X^1\leq x)\geq Pr(X^2\leq x)$ for all $x\in(-\infty, \infty)$.

\emph{Lemma 6 (Usual Stochastic Order Inequality \cite{Shaked2007} ):} If $X^1\leq_{\textrm{st}} X^2$, then $\mathbb{E}[X^1]\leq\mathbb{E}[X^2]$.

Combining \emph{Lemma 2}, we can see that $Pr(X\leq x)=0$ for $x<\textrm{Tr}[\overline{P}]$, $Pr(X\leq x)=\sum_{j=0}^i\frac1T\sum_{k=1}^Tp_{j,k}(\phi^1)$ for $\textrm{Tr}[h^{i}(\overline{P})]\leq x<\textrm{Tr}[h^{i+1}(\overline{P})], i=0,1,\ldots,T-1$, and $Pr(X\leq x)=1$ for $x\geq \textrm{Tr}[h^{T}(\overline{P})]$, to which $Pr(Y\leq x)$ has the similar structure. From \emph{Lemma 6}, the sign judgement of equation \eqref{JAD} can be recast as the check of whether there exists the relationship in \emph{Definition 2} between the induced random variables $X$ and $Y$. Based on the above analysis, we focus on the term $V_i=\sum_{j=0}^iF_j$, for $i=0,\ldots,T-1$ in the sequel, where $F_j=\sum_{k=1}^T[p_{j,k}(\phi^1)-p_{j,k}(\phi^2)]$.

Suppose $\theta\in\mathcal{S}_n$ and $\beta\in\mathcal{S}_{n+1}$. Note that there are $n$ times attacks in $\phi^1$ and $n+1$ times attacks in $\phi^2$.  Hence, the packet dropout probability in the presence of attack is $\theta$ under $\phi^1$ and $\beta$ under $\phi^2$.  And thereby, $V_i$ refers to $V_i(m,n,s,\theta,\beta)$. How to calculate $V_i$ is shown in the following lemma.

\emph{Lemma 7:} For simplicity, we write $V_i(m,n,s,\theta,\beta)$ as $V_i$ in this lemma. Let $N=n+s$ and $s< n$. There holds
\begin{align}\label{HI1}
V_i=&\,\,(n-i+1)\beta^{i+1}-(n-i)\theta^{i+1}-\alpha^{i+1}\nonumber\\
&+2\sum_{j=1}^i\alpha^j(\beta^{i-j+1}-\theta^{i-j+1}),
\end{align}
for $i= 0, \ldots, s-1$, and
\begin{align}\label{HI2}
V_i=(T-i)(\alpha^{i-n}\beta^{n+1}-\alpha^{i-n+1}\theta^n),
\end{align}
for $i= N, \ldots, T-1$.

When $m=s$, $V_i$ is given by
\begin{align}\label{HI3}
V_i=&\,\,(n-i+1)\beta^{i+1}-(n-i)\theta^{i+1}+\alpha^s\beta^{i-s+1}\nonumber\\
&-2\alpha^s\theta^{i-s+1}+2\sum_{j=1}^{s-1}\alpha^j(\beta^{i-j+1}-\theta^{i-j+1}),
\end{align}
for $i= s, \ldots, n$, and
\begin{align}\label{HI4}
V_i=&\,\,\alpha^s\beta^{i-s+1}-2\alpha^s\theta^{i-s+1}\nonumber\\
&+2\sum_{j=i-n+1}^{s-1}\alpha^j
(\beta^{i-j+1}-\theta^{i-j+1})\nonumber\\
&+(i-n+1)\alpha^{i-n}\beta^{n+1}-(i-n)\alpha^{i-n+1}\theta^n,
\end{align}
for $i= n+1, \ldots, N-1$.

When $m=s-1$, $V_i$ takes the form of
\begin{align}\label{HI5}
V_i=&\,\,(n-i+1)\beta^{i+1}-(n-i)\theta^{i+1}-\alpha^s\theta^{i+1-s}\nonumber\\
&+2\sum_{j=1}^{s-1}\alpha^j(\beta^{i-j+1}-\theta^{i-j+1}),
\end{align}
for $i= s, \ldots, n$, and
\begin{align}\label{HI6}
V_i=&\,\,-\alpha^s\theta^{i-s+1}+2\sum_{j=i-n+1}^{s-1}\alpha^j
(\beta^{i-j+1}-\theta^{i-j+1})\nonumber\\
&+(i-n+1)\alpha^{i-n}\beta^{n+1}-(i-n)\alpha^{i-n+1}\theta^n,
\end{align}
 for $i= n+1, \ldots, N-1$.
\begin{IEEEproof}
 See the Appendix.
\end{IEEEproof}

Next we present some results on the sign of $V_i$.

\emph{Lemma 8:} Let $\overline{n}=\lfloor\Delta/\underline{\delta}\rfloor$ and $\underline{n}=\lfloor\Delta/\overline{\delta}\rfloor$.
For any attack times $n$ with $\underline{n}\leq n\leq\overline{n}-1$, we have $V_i\geq 0$, for $i=0,\ldots,T-1$, if the following conditions are satisfied:
\begin{enumerate}
  \item $2\alpha\overline{\beta}^{\overline{n}-1}-\underline{\beta}^{\overline{n}}\leq 0$,
  \item $V_i(m,n,s,\overline{\beta},\underline{\beta})\geq0$, for $i=0,\ldots,T-1$, $n=\overline{n}-1$, and $n=\underline{n}$, where $\alpha$, $\overline{\beta}$ and $\underline{\beta}$ are the same with \emph{Proposition 1}, $m+s=T-n$, and $m=s$ or $m=s-1$.
\end{enumerate}
\begin{IEEEproof}
 See the Appendix.
\end{IEEEproof}

\emph{Remark 2:} \emph{Lemma 7} presents the expression of $V_i$ for $s<n$ which matters in the proof of \emph{Lemma 8}. As shown in the end of the proof of \emph{Lemma 8}, the case when $s\geq n$ is totally similar and is omitted for brevity. And note that it is beneficial for the proof of \emph{Lemma 8} but inconvenient for the verification of the conditions of \emph{Lemma 8} to employ the expression of $V_i$. It is easier to adopt numerical methods to compute $V_i$ based on its definition $V_i=\sum_{j=0}^iF_j$ when verifying the proposed sufficient conditions.

Then similar to \emph{Proposition 1}, we present the following proposition to show the effect of attack times on the remote estimation performance measured by the average error.

\emph{Proposition 2:} Let $\overline{n}=\lfloor\Delta/\underline{\delta}
\rfloor$.
Suppose $\theta\in\mathcal{S}_n$ and $\beta\in\mathcal{S}_{n+1}$, where $\underline{n}\leq n\leq\overline{n}-1$.
If the conditions in \emph{Lemma 8} are met, then we have
\begin{align*}
\textrm{Tr}(J_{max}^A)(\theta)\leq \textrm{Tr}(J_{max}^A)(\beta).
\end{align*}
\begin{IEEEproof}
We have $Pr(X\leq x)\geq Pr(Y\leq x)$ for all $x\in(-\infty, \infty)$ since $V_i\geq 0$, for $i=0,\ldots,T-1$. And thereby there holds $X\leq_{\textrm{st}} Y$, which leads to, due to \eqref{JAD} and \emph{Lemma 6}, $\textrm{Tr}(J_{max}^A)(\theta)\leq \textrm{Tr}(J_{max}^A)(\beta)$. The proof is completed.
\end{IEEEproof}

\subsection{Closed-form Solution}
The following lemma is needed before presenting the main result.

\emph{Lemma 9:} If $\beta_1, \beta_2\in \mathcal{S}_n$, and
$\beta_1>\beta_2$, then the attack strategy in \emph{Lemma 5} is optimal for $\beta_1$ and $\beta_2$. And there holds
\begin{align*}
\textrm{Tr}(J_{max}^A)(\beta_1)>\textrm{Tr}(J_{max}^A)(\beta_2).
\end{align*}
\begin{IEEEproof}
A direct result from \emph{Lemma 5}.
\end{IEEEproof}

Now the main conclusion is provided as follows.

\emph{Theorem 2:} Let $\overline{n}=\lfloor\Delta/\underline{\delta}\rfloor$ and $\underline{n}=\lfloor\Delta/\overline{\delta}\rfloor$. If conditions 1) and 2) in \emph{Lemma 8} are satisfied, then the optimal attack power level for the average error is
\begin{align}\label{the2}
\delta^{a}_*= \left\{ \begin{array}{ll}
\textrm{arg max $\{\delta^{a}|\lfloor\Delta/\delta^{a}\rfloor=\overline{n}\}$} ,& \textrm{for $\underline{n}<\overline{n}$}, \\
\overline{\delta} ,& \textrm{for $\underline{n}=\overline{n}$},
\end{array} \right.
\end{align}
the solution to \emph{Problem 1} for the average error is
$$\delta_{\ast}=(\underbrace{0,0,\ldots,0}_{m\, {\rm times}},\underbrace{\delta^{a}_*,\delta^{a}_*,\ldots,\delta^{a}_*}_{\overline{n}\, {\rm times}},\underbrace{0,0,\ldots,0}_{s\, {\rm times}}),$$
where $m+s=T-\overline{n}$, and $\vert{m-s}\vert\leq1$, i.e., $m=s$ or $\vert{m-s}\vert=1$, and the corresponding average expected estimation error can be calculated based on \emph{Lemma 5}.
\begin{IEEEproof}
To maximize the average error, in light of \emph{Lemma 5} and \emph{Proposition 2}, the attacker will block the channel $\overline{n}$ times. From \emph{Lemma 9}, the attacker will employ the larger power level. Hence, the conclusion in \emph{Theorem 2} holds.
\end{IEEEproof}

To maximize the average error, from \emph{Theorem 2}, the attacker should adopt the same action with the one for the terminal error except that the attacker should consecutively jam the channel in the middle of the considered time horizon when the proposed sufficient conditions hold.

\subsection{Exhaustion Search Method}
When the conditions in \emph{Lemma 8} are not met,  \eqref{the2} may not be the optimal one. Similar to the scenario for the terminal error, from \emph{Lemma 9}, we can transform \emph{Problem 1} for the average error into an equivalent problem:

\emph{Problem 5:}
\begin{eqnarray*}
& & \max \quad \textrm{Tr}[J_{max}^A(\delta)] \\
& & s.t. \quad (\delta^a,M)\in \Omega,
\end{eqnarray*}
where, $\delta=(\underbrace{0,0,\ldots,0}_{m\, {\rm times}},\underbrace{\delta^{a},\delta^{a},\ldots,\delta^{a}}_{\overline{n}\, {\rm times}},\underbrace{0,0,\ldots,0}_{s\, {\rm times}})$, $m+s=T-\overline{n}$, $\vert{m-s}\vert\leq1$, i.e., $m=s$ or $\vert{m-s}\vert=1$, $\Omega=\Big\{(f(n),n), n=\underline{n},\underline{n}+1,\cdots,\overline{n}\Big\}$, and $f(n)=\textrm{arg max}\{x|\lfloor\Delta/x\rfloor=n,\underline{\delta}\leq x\leq \overline{\delta}\}$.

Similarly, an exhaustion search method can be adopted to solve \emph{Problem 5}, with no more than $T$ steps required to obtain the solution.

\section{Optimal Dynamic Attack Energy Management}
The case when the attacker jams the channel with the constant energy is discussed in the previous sections. In this section we assume that the adversary has the capabilities of intercepting the real-time ACK information which indicates the arrival of packets or not, and dynamically adjusting the jamming energy based on the ACK signal. Then the optimal dynamic attack energy management problem arises.

More specifically, different from the static case where the attack strategy is limited to the set $\Theta=\{0,\delta^a\}^T$, here the attack power $\delta_k$ is dependent on the error $P_{k-1}$ and the available power $E_k$, i.e., the adversary has the attack policy $\delta=\{\delta_{1}, \delta_{2},\ldots,\delta_{T}\}$ with $\delta_k=\delta_k(P_{k-1},E_k)$. And similar to \cite{Shi2012,Peng2017,Zhang2018}, assume that the attacker has the discrete set of available power level $\Phi=\{0,\underline{\delta},e_1,e_2,\ldots,e_n,\overline{\delta}\}$ with $n$ finite and $0<\underline{\delta}<e_1<\cdots<e_n<\overline{\delta}$. Then the optimal dynamic attack energy management problem is formulated as follows.

\emph{Problem 6:}
\begin{eqnarray*}
& & \max_{\delta} \quad \textrm{Tr}[J^e(\delta)] \\
& & s.t. \quad\sum_{k=1}^{T}\delta_{k}\leq \Delta,\\
& & \qquad \,\,\,\delta_k\in\Phi.
\end{eqnarray*}

To solve \emph{Problem 6}, we formulate it as a finite Markov decision problem based on the Markov decision process (MDP) $\{T+1,\mathbb{S}, \mathbb{A}, Pr(\cdot|\cdot,\cdot),r_k(\cdot,\cdot)\}$ with the initial state $s_1=(\overline{P},\Delta)$. More specifically, $T+1$ is the considered time horizon.\footnote{We consider $T+1$ instead of $T$ since the error $P_T$ and the available power level $E_{T+1}$ at time $T+1$, i.e., the state $s_{T+1}$, is obtained after the decision is made at time $T$. But no decision is made at time $T+1$ and the process ceases.} The state space is defined as $\mathbb{S}=S_P\times S_E$, where $S_P=\{\overline{P},h(\overline{P}),h^{2}(\overline{P}),\ldots\}$ is a countable set associated with the estimation error covariance and $S_E=\{E^1,E^2,\ldots,E^r\}$ is the set of all the possible available power level at each time instant. And the state at time $k$ is defined as $s_k=(P_{k-1},E_k)\in \mathbb{S}$. The iteration processes of $P_k$ and $E_k$ are \eqref{Pk} and $E_k=E_{k-1}-\delta_{k-1}$, respectively.

Then we define the action space as $\mathbb{A}=\Phi$. For a given state $s=(h^{i}(\overline{P}),E^j)$, the set of allowable actions in state $s$ is $\mathbb{A}_s=\Phi\bigcap [0,E^j]$, and thereby, at time $k$, the attacker can choose an action $a_k$ from $\mathbb{A}_{s_k}$ for $k=1,\ldots,T$.

We further present the probability $Pr(s_{k+1}|s_k,a_k)$ that the state changes from $s_k=(h^{i}(\overline{P}),E^j)$ to $s_{k+1}$ with action $a_k$ taken at time $k$ for $k=1,\ldots,T$. From the iteration process of $s_k$, we have
\begin{align}\label{Prt}
&Pr(s_{k+1}|s_k,a_k)\nonumber\\
&= \left\{ \begin{array}{ll}
 \beta_k,& \textrm{if}\, s_{k+1}=(h^{i+1}(\overline{P}),E^j-a_k),\\
1-\beta_k,& \textrm{if}\, s_{k+1}=(\overline{P},E^j-a_k),\\
0 ,& \textrm{otherwise},
\end{array} \right.
\end{align}
where $i=0,1,2,\ldots$, $j=1,\ldots,r$ and the dropout probability $\beta_k$ can be obtained from \eqref{pk}--\eqref{pksinr}.

The one-stage reward function at time $k$ is defined as $r_k(s_k,a_k)$ for $k=1,\ldots,T+1$. Note that $r_{T+1}=r_{T+1}(s_{T+1})=0$ since no decision is made at time $T+1$ and thereby no reward is provided. And except the first time instant $k=1$, $r_k$ is random due to the randomness of $s_k$. The explicit expression of $r_k$ will be presented in the sequel.

\begin{algorithm}
\caption{Backward Induction Algorithm}
\label{algorithm1}
\KwIn{$T;\mathbb{S}^k;\mathbb{A}_s;Pr;r_k;s_1=(\overline{P},\Delta)$}
\KwOut{Maximum total reward $R_{T+1}^{\delta^*}(s_1)=u_{1}^*(s_1)$;
Optimal deterministic Markovian policy $\delta^*=\{\delta_{1}^*, \delta_{2}^*,\ldots,\delta_{T}^*\}$}
Set $k=T+1$ and $u_{T+1}^*(s)=r_{T+1}(s)$ for all $s\in\mathbb{S}^k$.\\
\While {$k>1$}
{
Let $k=k-1$ and compute $u_{k}^*(s)$ for each $s\in\mathbb{S}^k$ by
$$u_{k}^*(s)=\max_{a\in\mathbb{A}_s}\Big\{r_k(s,a)+\sum_{j\in\mathbb{S}^k}Pr(j|s,a)u_{k+1}^*(j)\Big\}.$$
Set for each $s\in\mathbb{S}^k$
$$\mathbb{A}_{s,k}^*=\mathop{\argmax}_{a\in\mathbb{A}_s}\Big\{r_k(s,a)+\sum_{j\in\mathbb{S}^k}Pr(j|s,a)u_{k+1}^*(j)
\Big\}.$$
Let $\delta_k^*(s)\in \mathbb{A}_{s,k}^*$ for each $s\in\mathbb{S}^k$.
}
\end{algorithm}

The attack policy $\delta=\{\delta_{1}, \delta_{2},\ldots,\delta_{T}\}$ with $\delta_k=\delta_k(P_{k-1},E_k)$ is a deterministic Markovian policy for the above MDP\cite{Puterman2005}. The usual optimality criteria\cite{Puterman2005} for the above finite MDP with the initial state $s_1$ and the adopted policy $\delta$ is the expected total reward over the time horizon $T+1$ which is defined by
$$R_{T+1}^\delta(s_1)=\mathbb{E}_{s_1}^\delta\Big[\sum_{k=1}^Tr_k(s_k,a_k)+r_{T+1}(s_{T+1})\Big].$$

Based on $R_{T+1}^\delta(s_1)$, we can obtain the terminal error $J^T$ by defining $r_k(\cdot,\cdot)$ as $r_k(s,a)=0$ for $k=1,\ldots,T-1$ and $r_k(s,a)=\textrm{Tr}(\mathbb{E}[P_T])$ for $k=T$, and $T\cdot J^A$ by defining $r_k(s,a)=\textrm{Tr}(\mathbb{E}[P_k])$ for $k=1,\ldots,T$ when the sample of $s_k$ is $s$ and the sample of $a_k$ is $a$, where $\mathbb{E}[P_k]=(1-\beta_k(a))\overline{P}+\beta_k(a)h(P_{k-1}(s))$.

For a given sample of $s_t$, $s$, define $u_t^\delta$ for $t<T+1$ by
$$u_t^\delta(s)=\mathbb{E}_{s}^\delta\Big[\sum_{k=t}^Tr_k(s_k,a_k)+r_{T+1}(s_{T+1})\Big],$$
from which it is easy to see that $R_{T+1}^\delta(s)=u_1^\delta(s)$, and let $u_{T+1}^\delta(s)=r_{T+1}(s)$. Further let $\mathbb{S}^k=S_P^k\times S_E$, where $S_P^k=\{\overline{P},h(\overline{P}),\ldots,h^{k}(\overline{P})\}$.
Then according to \cite{Puterman2005}, \emph{Problem 6} for both indexes can be solved by the backward induction algorithm (Algorithm 1) respectively through inputting the corresponding $r_k$.
Note that in Algorithm 1, $\mathbb{S}^k$ is used instead of $\mathbb{S}$. This is from the fact that $S_P^k$ includes all the possible value of $P_k$ due to \eqref{Pk} and the given initial state $s_1=(\overline{P},\Delta)$, and thereby there is no need to compute all $s\in \mathbb{S}$.
And we can see from Algorithm 1 that $\delta^*$ may be not unique which occurs if $\mathbb{A}_{s,k}^*$ contains more than one action for some $s$ and $k$. We just need to retain a single action from $\mathbb{A}_{s,k}^*$ at this time to acquire a particular optimal policy.

\section{Optimal Tradeoff Between Attack Energy Consumption and System Degradation}
The case when the attacker has the fixed total energy constraint $\Delta$ is discussed in the previous sections. It is well-known that more attack energy used leads to more system degradation but more energy consumption. The attacker may desire a tradeoff between energy expense and system degradation by decreasing the employed attack energy at the cost of weakening the attack effect. Then one question arises that how much energy the attacker should decrease to achieve the optimal tradeoff? To answer this question, here we propose the modified Markov decision problem based on Section V. The modified parts are presented as follows.

With no total energy restriction imposed, the state space in this section is defined as $\mathbb{\tilde{S}}=S_P$ and the state at time $k$ as $\tilde{s}_k=P_{k-1}\in \mathbb{\tilde{S}}$. The action space is still $\mathbb{\tilde{A}}=\Phi$ but for any given state $s$, the attacker can choose an action $a$ from $\mathbb{\tilde{A}}_s=\mathbb{\tilde{A}}$. And the one-stage reward function at time $k$ is $R(\tilde{s}_k,a_k)$. Corresponding to the terminal error $J^T$ and the average error $J^A$, we can respectively design that $R_k(s,a)=-\omega a$ for $k=1,\ldots,T-1$ and $r_k(s,a)=\textrm{Tr}(\mathbb{E}[P_T])-\omega a$ for $k=T$, and $R_k(s,a)=\textrm{Tr}(\mathbb{E}[P_k])-\omega a$ for $k=1,\ldots,T$ when the sample of $\tilde{s}_k$ is $s$ and the sample of $a_k$ is $a$, where $\mathbb{E}[P_k]=(1-\beta_k(a))\overline{P}+\beta_k(a)h(s)$ and $\omega>0$ is the weighting parameter. From the design of $R_k$, the modified objective is to maximize the attack effect and minimize the energy expense simultaneously. Similar objective function appears in \cite{Li2017,Adibi2007}. We call the modified Markov decision problem the optimal tradeoff problem between
attack energy and system degradation, which is based on the modified MDP $\{T+1,\mathbb{\tilde{S}}, \mathbb{\tilde{A}}, Pr(\cdot|\cdot,\cdot),R_k(\cdot,\cdot)\}$ with the initial state $\tilde{s}_1=\overline{P}$. And the corresponding $\tilde{R}_{T+1}^\delta(\tilde{s}_1)$ and $\tilde{u}_t^\delta(s)$ can be obtained by replacing $r_k$ and $s_1$ in $R_{T+1}^\delta(s_1)$ and $u_t^\delta(s)$ with $R_k$ and $\tilde{s}_1$, respectively.

Before presenting the algorithm that solves the proposed optimal tradeoff problem, we derive some structural results for the corresponding optimal policy. To do this, first the following definition is introduced.

\emph{Definition 3 (\cite{Puterman2005}):} Let $\hat{X}$ and $\hat{Y}$ be partially ordered sets and $g(x,y)$ a real-valued function on $\hat{X}\times \hat{Y}$. We say that $g$ is superadditive if for $x^+\geq x^-$ in $\hat{X}$ and $y^+\geq y^-$ in $\hat{Y}$, there holds
\begin{align*}
g(x^+,y^+)+g(x^-,y^-)\geq g(x^+,y^-)+g(x^-,y^+).
\end{align*}

Then the following theorem provides some structural results for the optimal policy $\delta_k^*$.

\emph{Theorem 3:} There exist optimal decision rules $\delta_k^*(s)$ which are nondecreasing in $s$ for $k=1,\ldots,T$.
\begin{IEEEproof}
See the Appendix.
\end{IEEEproof}
\emph{Theorem 3} is desirable since it helps reduce the search range when seeking the optimal policy. Specifically, the monotone backward induction algorithm (Algorithm 2) in which $\mathbb{\tilde{S}}^k=S_P^k$ is presented to solve the optimal tradeoff problem.
\begin{algorithm}
\caption{Monotone Backward Induction Algorithm}
\label{algorithm1}
\KwIn{$T;\mathbb{\tilde{S}}^k;\mathbb{\tilde{A}};Pr;R_k;\tilde{s}_1=\overline{P})$}
\KwOut{Maximum total reward $\tilde{R}_{T+1}^{\delta^*}(\tilde{s}_1)=\tilde{u}_{1}^*(\tilde{s}_1)$;
Optimal deterministic Markovian policy $\delta^*=\{\delta_{1}^*, \delta_{2}^*,\ldots,\delta_{T}^*\}$}
Set $k=T+1$ and $\tilde{u}_{T+1}^*(s)=R_{T+1}(s)$ for all $s\in\mathbb{\tilde{S}}^k$.\\
\While {$k>1$}
{
Let $k=k-1$. Set $s=\overline{P}$ and $\mathbb{\tilde{A}}_{\overline{P}}=\mathbb{\tilde{A}}$.\\
\While {$s<h^k(\overline{P})$}
{
Compute $\tilde{u}_{k}^*(s)$ by
$$\tilde{u}_{k}^*(s)=\max_{a\in\mathbb{\tilde{A}}_s}\Big\{R_k(s,a)+\sum_{j\in\mathbb{\tilde{S}}^k}Pr(j|s,a)
\tilde{u}_{k+1}^*(j)\Big\}.$$
Set
$$\mathbb{\tilde{A}}_{s,k}^*=\mathop{\argmax}_{a\in\mathbb{\tilde{A}}_s}\Big\{R_k(s,a)+\sum_{j\in
\mathbb{\tilde{S}}^k}Pr(j|s,a)\tilde{u}_{k+1}^*(j)\Big\}.$$
Let $\delta_k^*(s)\in \mathbb{\tilde{A}}_{s,k}^*$.
Set
$$\mathbb{\tilde{A}}_{h(s)}=\{a\in \mathbb{\tilde{A}}: a\geq \textrm{max}[a'\in \mathbb{\tilde{A}}_{s,k}^*]\}.$$
Let $s=h(s)$.
}
}
\end{algorithm}

\section{Simulations And Examples}
In this section we show the system performance under the proposed DoS attack with the optimal attack power level for the terminal error and the average error, respectively.  First, we evaluate the effects of attacks with different power levels when the conditions in \emph{Proposition 1} are satisfied  for $\alpha=0$ to verify that ours are more relaxed than the sufficient conditions in \cite{Zhang2018}. Then the optimal static attack energy management for the terminal error is obtained based on the exhaustion search method when the conditions in \emph{Proposition 1} do not hold for $\alpha$ which is obtained based on \eqref{pk}--\eqref{pksinr}. Similarly, simulation examples for the average error are provided in the sequel. We also solve the optimal dynamic energy management problem and the optimal tradeoff problem, respectively, based on Algorithm 1 and Algorithm 2, which implies that the optimal dynamic policy has better performance than the optimal static policy, and verifies that the optimal policy for the optimal tradeoff problem has the monotone structure. The system parameters $A$, $C$, $Q$ and $R$, are given in \emph{Example 1}.

\subsection{Closed-form Solution for Terminal Error}
In this subsection, we will adopt the same parameters with Fig. 6 in \cite{Zhang2018} to verify the relaxation of our proposed sufficient conditions. Let the packet dropout probability without attack be $\alpha=0$. And the packet dropout probability without attack $\alpha$ is obtained based on \eqref{pk}--\eqref{pksinr} in all the subsequent subsections. The sensor sends the data packet with the packet length $L=20$ to the remote estimator with the power $\delta^s=10$ through a wireless link with the channel gain  $G_s=1$ and the noise power $\sigma^2=2$.
The channel gain for the attacker is $G_a=1$. The maximal available power is $\Delta=200$. The
lower bound and upper bound of $\delta^a$, respectively, are $\underline{\delta}=20$ and $\overline{\delta}=50$. Note that the value of $T$ is not needed due to the expression of the terminal error in \eqref{Jmax} and the fact that $\alpha=0$.

In \cite{Zhang2018}, the authors employ the exhaustion search method to find the optimal attack level, since the system matrix $A$ is not normal. However, according to \emph{Example 1}, the condition 1) in \emph{Proposition 1} is satisfied. And it is easy to verify the satisfaction of the condition 2) in \emph{Proposition 1}. Therefore, the optimal attack level is $\underline{\delta}=20$ from \emph{Theorem 1}, which is also illustrated in Fig. 2.
\subsection{Solution for Terminal Error from Exhaustion Search Method}
All parameters are the same with \emph{Section VII. A} except that $\Delta=50$, $\underline{\delta}=2$, $\overline{\delta}=20$, and the considered time horizon is $T=30$. Then the condition 2) in \emph{Proposition 1} does not hold. Hence, $f(\overline{n})=\textrm{arg max}\{x|\lfloor\Delta/x\rfloor=\overline{n},\underline{\delta}
\leq x\leq \overline{\delta}\}$ may not be the optimal attack level, with $\overline{n}=\lfloor\Delta/\underline{\delta}\rfloor=25$. As shown in Fig. 3, the optimal attack level is $\delta^{a}_*=50/6$, with the optimal attack times 6 and the maximal trace of expected terminal error covariance 17.5865. And note that the conditions in \emph{Proposition 1} hold if $\underline{\delta}=10$ and $\overline{\delta}=20$, i.e., $10$ is the optimal attack level if $\underline{\delta}=10$ and $\overline{\delta}=20$, which can be also seen from Fig. 3.

\begin{figure}
	\centering
	\includegraphics[width=1\linewidth]{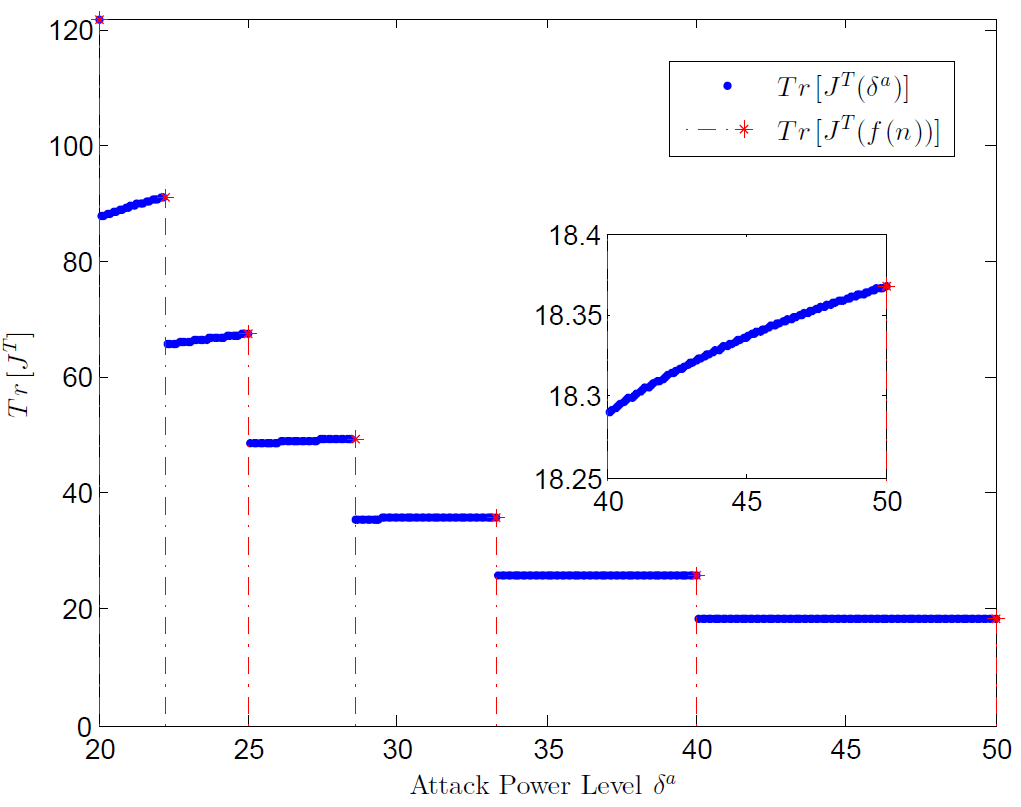}\\
	\caption{$Tr[J^T]$ under different attack power levels when the sufficient conditions in \emph{Proposition 1} hold.}
\end{figure}

\begin{figure}
  \centering
  \includegraphics[width=1\linewidth]{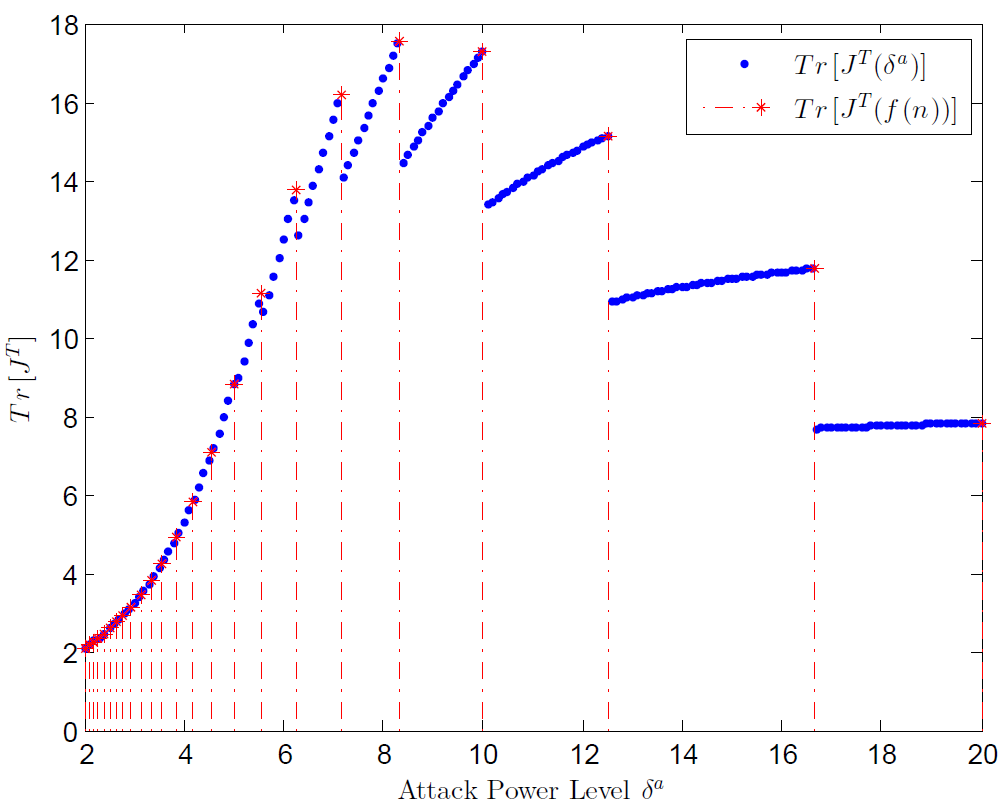}\\
  \caption{$Tr[J^T]$ under different attack power levels when the sufficient conditions in \emph{Proposition 1} are not satisfied.}
\end{figure}

\subsection{Closed-form Solution for Average Error}
In this subsection, we will adopt the same parameters with \emph{Section VII. A} except that the considered time horizon is $T=15$. It is easy to verify the satisfaction of the condition 1) and 2) in \emph{Lemma 8}. Therefore, the optimal attack level is $\underline{\delta}=20$ from \emph{Theorem 2}, which is also illustrated in Fig. 4.

\subsection{Solution for Average Error from Exhaustion Search Method}
All parameters are the same with \emph{Section VII. B}. Then the conditions in \emph{Lemma 8} do not hold. Hence, $f(\overline{n})=\textrm{arg max}\{x|\lfloor\Delta/x\rfloor=\overline{n},\underline{\delta}
\leq x\leq \overline{\delta}\}$ may not be the optimal attack level, with $\overline{n}=\lfloor\Delta/\underline{\delta}\rfloor=25$. As shown in Fig. 5, the optimal attack level is $\delta^{a}_*=50/8$, with the optimal attack times 8 and the maximal trace of expected average error covariance 3.2435.

\begin{figure}
  \centering
  \includegraphics[width=1\linewidth]{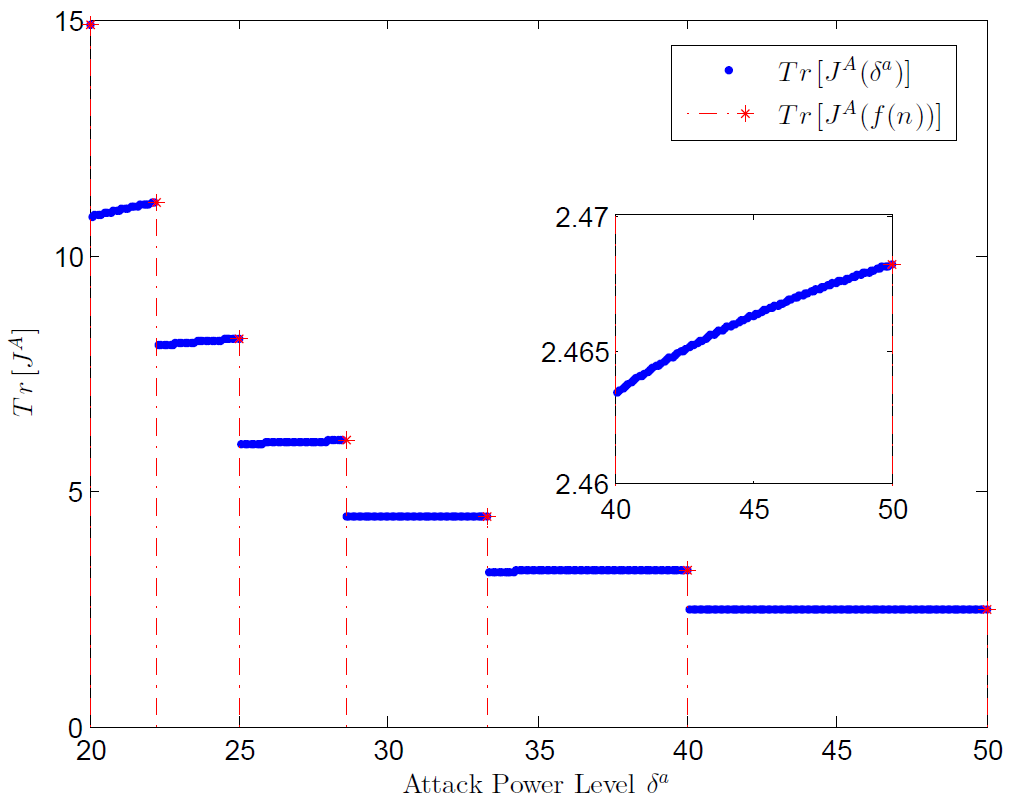}\\
  \caption{$Tr[J^A]$ under different attack power levels when the sufficient conditions in \emph{Lemma 8} hold.}
\end{figure}

\begin{figure}
  \centering
  \includegraphics[width=1\linewidth]{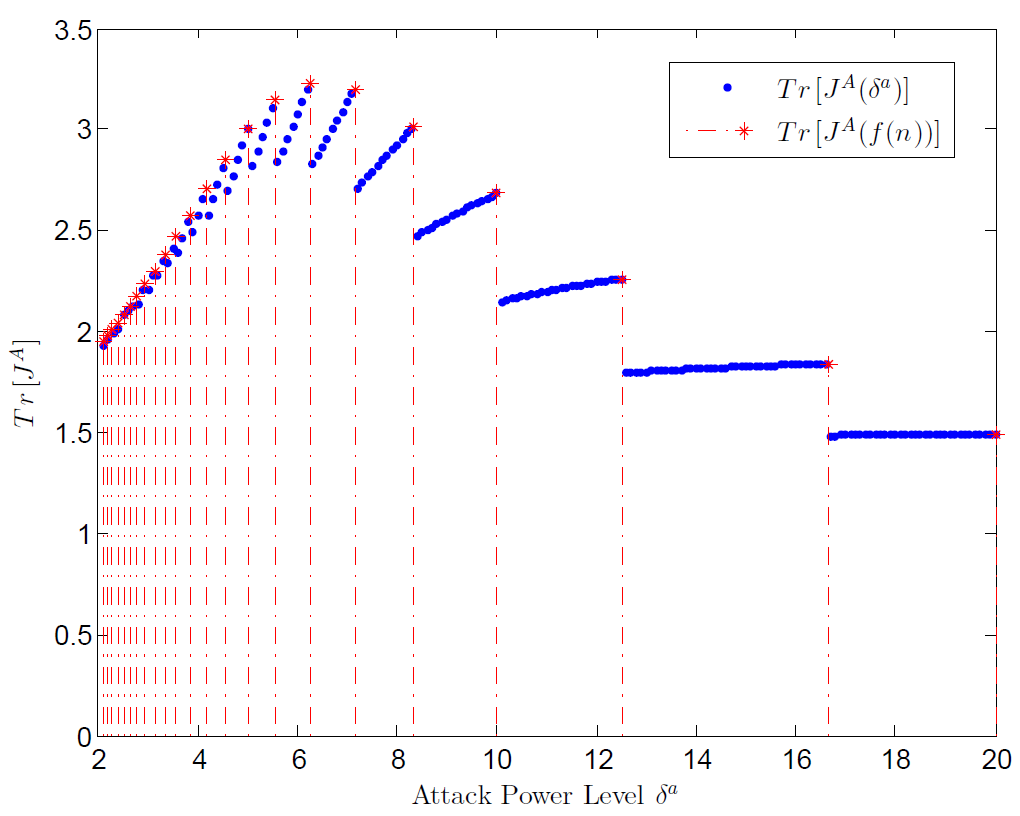}\\
  \caption{$Tr[J^A]$ under different attack power levels when the sufficient conditions in \emph{Lemma 8} are not satisfied.}
\end{figure}

\subsection{Optimal Dynamic Attack Power Allocation with Given Energy Constraint}
In this subsection, we examine the proposed Algorithm 1. Take the average error for example. Let the available power level set be $\Phi=\{0,5,10,15\}$. Here for ease of simulation, set the corresponding probability set as $\{0.1,0.3,0.7,0.9\}$. The total energy constraint is $\Delta=60$ and the considered time horizon is $T=5$. Then based on Algorithm 1, we can utilize the value iteration algorithm in \cite{Puterman2005} to find the optimal dynamic attack power policy which is described by Fig. 6. The arrow goes to the possible state at the next time step. Here $h^k$ refers to $h^k(\overline{P})$. And under the optimal policy, the maximum trace of average error, $R_{T+1}^{\delta^*}(s_1)=8.0256$, is achieved.

\begin{figure}
  \centering
  \includegraphics[width=0.95\linewidth]{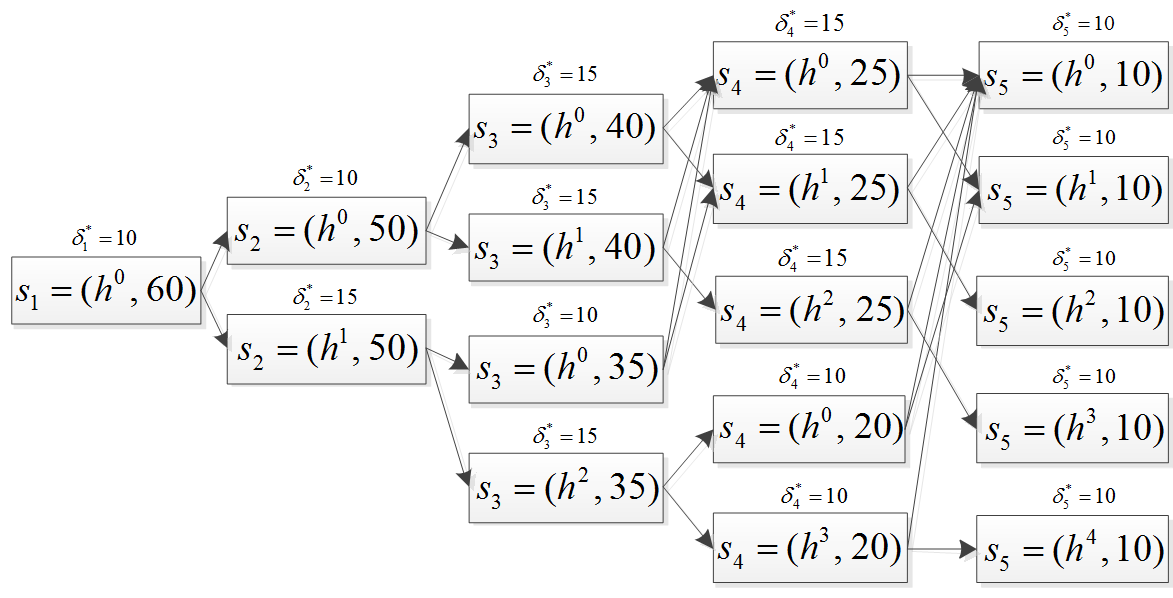}\\
  \caption{Optimal decision tree for the optimal dynamic attack management problem with the initial state $s_1=(\overline{P},60)$.}
\end{figure}

\begin{figure}
  \centering
  \includegraphics[width=1\linewidth]{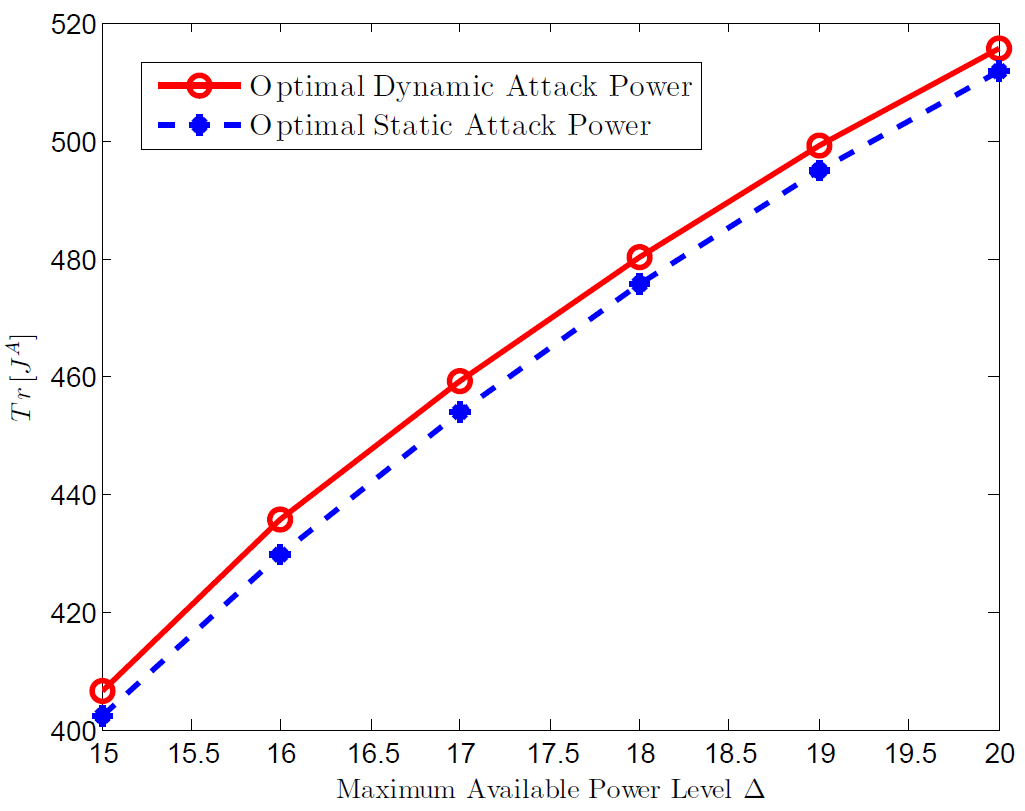}\\
  \caption{Maximum $Tr[J^A]$ with varying maximum attack power $\Delta$.}
\end{figure}

Then we compare the optimal static attack power policy and the optimal dynamic attack power policy proposed in Section IV and Section V, respectively, by changing the maximum available power $\Delta=15,16,\ldots,20$. For ease of illustration, we reset the system matrix as $A=\begin{bmatrix} 1.01 & 3 \\ 0 & 1 \end{bmatrix}$. Fix the available power level set as $\Phi=\{0,2,2.25,2.5,\ldots,9.75,10\}$ with $\underline{\delta}=2$ and $\overline{\delta}=10$. Set $T=7$, $L=20$, $\delta^s=2$, $G_s=1$, $G_a=1$ and $\sigma^2=0.5$. The corresponding packet dropout probability is calculated from \eqref{pk}--\eqref{pksinr}. The result of the above comparison is illustrated in Fig. 7, from which we can see that the optimal dynamic policy has better performance than the optimal static policy, i.e., the attacker can degenerate the remote estimation quality more severely if it has the ability of acquiring the real-time ACK information. And note that $\Phi$ is not the interval $[2,10]$. The attack effect can be further improved if the available power set includes more elements such as $\Phi=\{0,2,2.01,2.02,\ldots,9.99,10\}$.
\subsection{Dynamic Attack Power Allocation for Optimal Tradeoff Problem}
All parameters are the same with Fig. 6. Let the weighting coefficient be $\omega=0.35$. Similar to the above subsection, the optimal policy for the optimal tradeoff problem between system degradation and energy consumption is obtained by running the value iteration algorithm in \cite{Puterman2005}. Note that, Algorithm 2 is not employed and the search space of action is $\mathbb{\tilde{A}}$ instead of $\mathbb{\tilde{A}}_{h(s)}=\{a\in \mathbb{\tilde{A}}: a\geq \textrm{max}[a'\in \mathbb{\tilde{A}}_{s,k}^*]\}.$ This is to verify the monotonicity of the optimal policy in \emph{Theorem 3}. Then the optimal policy is shown in Fig. 8 from which we can see that for a given time step $k$, $\delta_k^*(s)$ is nondecreasing in $s$. This is consistent with the theoretical result in \emph{Theorem 3}. The maximum total reward corresponding to the average error is $\tilde{R}_{T+1}^{\delta^*}(\tilde{s}_1)=5.3325$.

\begin{figure}
  \centering
  \includegraphics[width=0.95\linewidth]{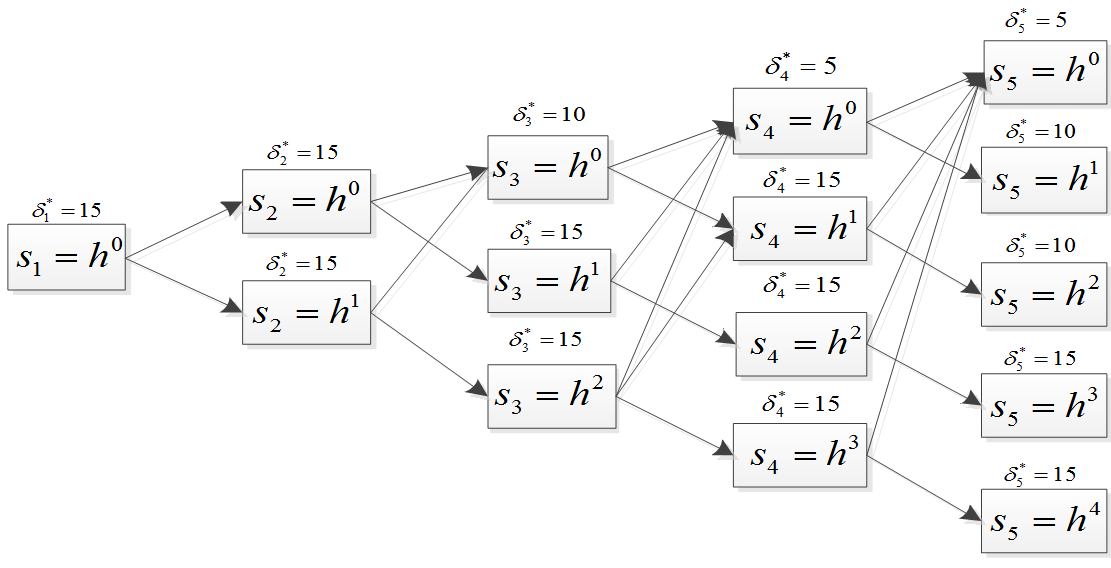}\\
  \caption{Optimal decision tree for the optimal tradeoff problem with the initial state $s_1=\overline{P}$.}
\end{figure}

\section{Conclusions}\label{sec:conclusion}
In this paper, a system in which remote state estimation is carried out was considered. We investigated how to allocate the constant attack power to maximize two kinds of indexes of system performance, \emph{Terminal error} and \emph{Average error}, respectively, at the remote estimator when an energy-constrained attacker launches a DoS attack against the SINR-based wireless channel. We proposed novel analysis approaches to derive some sufficient conditions for two kinds of indexes, respectively. An explicit solution to the issue of how much power should be adopted was attained for both two kinds of indexes if the system parameters meet the corresponding conditions. Further we demonstrated that our proposed conditions for \emph{Terminal error} are more relaxed than the one in the existing work. When the sufficient conditions fail to be satisfied, a feasible method was provided to find the optimal attack level for both two kinds of indexes. Then the case when the attacker could acquire the real-time ACK information and desires the time-varying attack power is studied and a MDP-based algorithm is designed to find the optimal dynamic attack power allocation. To optimize the tradeoff between system degradation and attack energy, the other MDP-based algorithm was further proposed based on which the optimal tradeoff can be found. And a monotone structure of the optimal policy was exploited such that the efficiency of the proposed algorithm can be improved dramatically. Finally, the effectiveness of the theory was verified by the numerical examples.

\begin{appendix}
In this section, we prove \emph{Proposition 1}, \emph{Lemma 7}, \emph{Lemma 8} and \emph{Theorem 3}. First the proof of \emph{Proposition 1} in detail is provided.

\begin{IEEEproof}[Proof of Proposition 1]
From the condition 2) in $\emph{Proposition 1}$, we can readily obtain that
\begin{align*}
\overline{\beta}^{\overline{n}-n-1}(\alpha\overline{\beta}^{n}
-\underline{\beta}^{n+1})\leq \alpha\overline{\beta}
  ^{\overline{n}-1}-\underline{\beta}^{\overline{n}}\leq 0.
\end{align*}

Since $0<\alpha<\underline{\beta}\leq\beta\leq\theta\leq
\overline{\beta}<1$, there holds
\begin{align}\label{cof}
\alpha\theta^n-\beta^{n+1}\leq\alpha\overline{\beta}^{n}
-\underline{\beta}^{n+1}\leq 0.
\end{align}

Let $\textrm{Tr}(J_{\theta,\beta}^T)=\textrm{Tr}(J^T)_{max}(\theta)- \textrm{Tr}(J^T)_{max}(\beta)$.
Then in light of \eqref{Jmaxn}, we have
\begin{align*}
\textrm{Tr}(J_{\theta,\beta}^T)
=&\,\textrm{Tr}\Big\{(\alpha\theta^n-\beta^{n+1})\sum_{i=n+1}^{T}\alpha^{i-n-1}H^i\\
&+\sum_{i=1}^{n}(\theta^i-\beta^i)H^i\Big\}.
\end{align*}

From \eqref{cof} and the condition 1) in $\emph{Proposition 1}$, the following inequality is true:
\begin{align}\label{Jine}
\textrm{Tr}(J_{\theta,\beta}^T)\leq&\,\textrm{Tr}\Big\{\big[\sum_{i=1}^{n}(\overline
{\beta}^i-\underline{\beta}^i)+(\alpha\overline{\beta}^{n}-
\underline{\beta}^{n+1})\sum_{i=0}^{T-n-1}\alpha^{i}\big]
\nonumber\\
&\times H^n\Big\}.
\end{align}

Due to \eqref{cof}, there holds
\begin{align*}
&[\alpha\overline{\beta}^{n}(1-\overline{\beta}^{\overline{n}
-n-1})+\underline{\beta}^{n+1}(\underline{\beta}^{\overline{n}
-n-1}-1)]\\
\leq&\,\underline{\beta}^{n+1}(\underline{\beta}^{\overline{n}
-n-1}-\overline{\beta}^{\overline{n}
-n-1})\leq 0,
\end{align*}
which causes
\begin{align*}
&(\alpha\overline{\beta}^{n}-\underline{\beta}^{n+1})\sum_{i=0}
^{T-n-1}\alpha^{i}-(\alpha\overline{\beta}^{\overline{n}-1}-
\underline{\beta}^{\overline{n}})\sum_{i=0}^{T-\overline{n}}
\alpha^{i}\\
=&\,[\alpha\overline{\beta}^{n}(1-\overline{\beta}^{\overline{n}
-n-1})+\underline{\beta}^{n+1}(\underline{\beta}^{\overline{n}
-n-1}-1)]\sum_{i=0}^{T-\overline{n}}\alpha^{i}\\
&+(\alpha\overline{\beta}^{n}-\underline{\beta}^{n+1})
\sum_{i=T-\overline{n}+1}
^{T-n-1}\alpha^{i}\leq 0.
\end{align*}
Hence, it is obtained that
\begin{align*}
&\sum_{i=1}^{n}(\overline
{\beta}^i-\underline{\beta}^i)+(\alpha\overline{\beta}^{n}-
\underline{\beta}^{n+1})\sum_{i=0}^{T-n-1}\alpha^{i}\\
\leq&\sum_{i=1}^{\overline{n}-1}(\overline{\beta}^i-\underline{\beta}
^i)+(\alpha\overline{\beta}^{\overline{n}-1}-\underline{\beta}
^{\overline{n}})\sum_{i=0}^{T-\overline{n}}\alpha^{i},
\end{align*}
which leads to, according to \eqref{Jine}, $\emph{Lemma 2}$, and the condition 2) in $\emph{Proposition 1}$,
\begin{align*}
\textrm{Tr}(J_{\theta,\beta}^T)\leq 0.
\end{align*}
The proof is completed.
\end{IEEEproof}

Then we present the proof of \emph{Lemma 7}.
\begin{IEEEproof}[Proof of Lemma 7]
Let $H(i,\lambda)=\sum_{k=1}^T\mathbb{E}[P_k(\lambda)]$ with $P_0=h^i(\overline{P})$, where $T$ is the length of the time horizon for an attack schedule $\lambda$.  Let $N=n+s$, $\psi^1=(\lambda^{n},\gamma^{s})$ and $\psi^2=(\lambda^{n+1},\gamma^{s-1})$.
From the structure of $\phi^1=(\gamma^{m},\lambda^{n},\gamma^{s})$, $\phi^2=(\gamma^{m},\lambda^{n+1},\gamma^{s-1})$ and $V_i=\sum_{j=0}^iF_j$, where $F_i=\sum_{k=1}^T[p_{i,k}(\phi^1)-p_{i,k}(\phi^2)]$, we can obtain
\begin{align*}
V_i(m,n,s,\theta,\beta)=&\sum_{j=0}^i\sum_{k=1}^T(p_{j,k}(\phi^1)-p_{j,k}(\phi^2))\nonumber\\
=&\sum_{j=0}^i\sum_{k=m+1}^T(p_{j,k}(\phi^1)-p_{j,k}(\phi^2)),
\end{align*}
and
\begin{align*}
&J_{max}^A(\phi^1)-J^A_{max}(\phi^2)\nonumber\\
=&\frac1T\sum_{k=1}^T\{\mathbb{E}[P_k(\phi^1)]-\mathbb{E}[P_k(\phi^2)]\}\nonumber\\
=&\frac1T\sum_{k=m+1}^T\{\mathbb{E}[P_k(\phi^1)]-\mathbb{E}[P_k(\phi^2)]\}
\nonumber\\
=&\frac1T\sum_{i=0}^{T}h^{i}(\overline{P})\sum_{k=m+1}^T
(p_{i,k}(\phi^1)-p_{i,k}(\phi^2)),
\end{align*}
where,
\begin{align*}
\sum_{k=m+1}^T\mathbb{E}[P_k(\phi^1)]=&\sum_{i=0}^{T}h^{i}(\overline{P})
\sum_{k=m+1}^Tp_{i,k}(\phi^1)\nonumber\\
=&\sum_{j=0}^{m-1}(1-\alpha)\alpha^jH(j,\psi^1)+\alpha^mH(m,\psi^1).
\end{align*}

According to the equation
\begin{align*}
H(j,\psi^1)=H(0,\psi^1)+\sum_{i=1}^Np_{i,i}[h^{i+j}(\overline{P})-h^i(\overline{P})],
\end{align*}
we can further obtain
\begin{align}\label{Vm}
&\sum_{k=m+1}^T\mathbb{E}[P_k(\phi^1)]\nonumber\\
=&H(0,\psi^1)+\sum_{j=0}^{m-1}(1-\alpha)\alpha^j\sum_{i=1}^Np_{i,i}[h^{i+j}(\overline{P})-
h^i(\overline{P})]\nonumber\\
&+\alpha^m\sum_{i=1}^Np_{i,i}[h^{i+m}(\overline{P})-h^i(\overline{P})]\nonumber\\
=&\, H(0,\psi^1)+\sum_{i=2}^m h^i(\overline{P})[\sum_{j=1}^{i-1}\alpha^{i-j}p_{j,j}-\sum_{j=1}^i\alpha^{i-j+1}p_{j,j}]
\nonumber\\
&+\sum_{i=m+1}^N h^i(\overline{P})[\sum_{j=i-m}^{i-1}\alpha^{i-j}p_{j,j}-\sum_{j=i-m+1}^i\alpha^{i-j+1}p_{j,j}]
\nonumber\\
&+\sum_{i=N+1}^{T-1} h^i(\overline{P})[\sum_{j=i-m}^N\alpha^{i-j}p_{j,j}-\sum_{j=i-m+1}^N\alpha^{i-j+1}p_{j,j}]
\nonumber\\
&+h^T(\overline{P})\alpha^mp_{N,N}-h(\overline{P})\alpha p_{1,1},
\end{align}
where, $p_{i,i}=p_{i,i}(\psi^1)$ and
\begin{align}\label{V0}
&\,\,H(0,\psi^1)\nonumber\\
=&\,\,\overline{P}[n(1-\theta)+s(1-\alpha)]+h^N(\overline{P})\theta^n\alpha^s
\nonumber\\
&+\sum_{i=1}^s h^i(\overline{P})\big[\theta^i+(n-i)(1-\theta)\theta^i\nonumber\\
&+\sum_{j=1}^i(1-\theta)\theta^{i-j}\alpha^{j}+(s-i)(1-\alpha)\alpha^i\big]\nonumber\\
&+\sum_{i=n+1}^{N-1}h^i(\overline{P})\big[\theta^n\alpha^{i-n}
+\sum_{j=1}^{N-i}(1-\theta)\theta^{n-j}\alpha^{i-n+j}\big]\nonumber\\
&+\sum_{i=s+1}^nh^i(\overline{P})\big[\theta^i+(n-i)(1-\theta)\theta^i\nonumber\\
&+\sum_{j=1}^s(1-\theta)\theta^{i-j}\alpha^{j}\big].
\end{align}

Before proceeding further, one equation is presented to facilitate the analysis:
\begin{align}\label{EQ1}
&\sum_{i=t_1}^{t_2}\sum_{j=t_3}^{i+t_3-t_1}(1-\theta)\theta^{i-j}
\alpha^j\nonumber\\
=&\sum_{j=t_3}^{t_2+t_3-t_1}\sum_{i=j+t_1-t_3}^{t_2}
(1-\theta)\theta^{i-j}\alpha^j\nonumber\\
=&\sum_{j=t_3}^{t_2+t_3-t_1}\theta^{t_1-t_3}
(1-\theta^{t_2-t_1+t_3+1-j})\alpha^j.
\end{align}

Based on \eqref{Vm} and \eqref{V0}, for $i= 0, \ldots, s-1$,  there holds
\begin{align}
&\sum_{j=0}^i\sum_{k=m+1}^Tp_{j,k}(\phi^1)\nonumber\\
=&\,\,\sum_{j=0}^i[(n-j)(1-\theta)\theta^j+(s-j)(1-\alpha)\alpha^j]+\sum_{j=1}^i
\theta^j\nonumber\\
&+2\sum_{j=2}^i\sum_{t=1}^j(1-\theta)\theta^{j-t}\alpha^{t}-\sum_{j=2}^i\alpha^j
+\alpha(1-2\theta),\nonumber
\end{align}
which, combining \eqref{EQ1} and the equation

\begin{align}
\sum_{j=0}^i(n-j)(1-\theta)\theta^j=n-(n-i)\theta^{i+1}-\sum_{j=1}^i\theta^j,\nonumber
\end{align}
leads to
\begin{align*}
V_i=&\,\,(n-i+1)\beta^{i+1}-(n-i)\theta^{i+1}-\alpha^{i+1}\nonumber\\
&+2\sum_{j=1}^i\alpha^j(\beta^{i-j+1}-\theta^{i-j+1}),
\end{align*}
i.e., equation \eqref{HI1}.

Similarly, equations \eqref{HI2}--\eqref{HI6} can be derived from \eqref{Vm} and \eqref{V0}.
The proof is completed.
\end{IEEEproof}
Next, the proof of \emph{Lemma 8} is given as follows.

\begin{IEEEproof}[Proof of Lemma 8]
It is hard to compare $V_i's$ with different attack times since different attack times $n$ bring different packet dropout probabilities under attack. To proceed to the next analytic step, we find the lower bound of $V_i(m,n,s,\theta,\beta)$. From \eqref{HI1}--\eqref{HI6}, there holds $V_i(m,n,s,\theta,\beta)\geq V_i(m,n,s,\overline{\beta},\underline{\beta})$, since each term with $\theta$ is less than 0 and each term with $\beta$ is positive. Next we focus on the comparison between the obtained lower bounds of $V_i(m,n,s,\theta,\beta)$, i.e., $V_i(m,n,s,\overline{\beta},\underline{\beta})$. More specifically, we investigate how $V_i(m,n,s,\overline{\beta},\underline{\beta})$ varies with $n$. And for simplicity of description, we write $V_i(m,n,s,\overline{\beta},\underline{\beta})$ as $V_i(m,n,s)$ throughout the following derivation.

For ease of understanding, here we rewrite $\phi^1=(\gamma^{m},\lambda^{n},\gamma^{s})$, and $\phi^2=(\gamma^{m},\lambda^{n+1},\gamma^{s-1})$. It is easy to see from \emph{Lemma 7} that $\phi^1$ has two kinds of structures, $(\gamma^{s},\lambda^{n},\gamma^{s})$ and $(\gamma^{s-1},\lambda^{n},\gamma^{s})$. According to \emph{Lemma 7}, the expression of $V_i$ is dependent on the structure of $\phi^1$. Let $m=d$, $n=k$ and $s=d$. And we fix the length of the time horizon as $T=d+k+d$. Then $\phi^1$ has the first structure $(\gamma^{s},\lambda^{n},\gamma^{s})$, and the corresponding $\phi^2$ is $(\gamma^{d},\lambda^{k+1},\gamma^{d-1})$ which, from \emph{Lemma 5}, leads to the same average error with  $(\gamma^{d-1},\lambda^{k+1},\gamma^{d})$. When $n=k+1$, $\phi^1=(\gamma^{d-1},\lambda^{k+1},\gamma^{d})$ has the second structure $(\gamma^{s-1},\lambda^{n},\gamma^{s})$ since $T=d+k+d$, and the corresponding $\phi^2$ is $(\gamma^{d-1},\lambda^{k+2},\gamma^{d-1})$. Similarly, when $n=k+2$, $\phi^1=(\gamma^{d-1},\lambda^{k+2},\gamma^{d-1})$, again, has the first structure $(\gamma^{s},\lambda^{n},\gamma^{s})$, and the corresponding $\phi^2$ is $(\gamma^{d-1},\lambda^{k+3},\gamma^{d-2})$. Hence, different expressions of $V_i$ should be adopted as the attack times $n$ increases. In the following derivation, we first focus on the comparison between $V_i(d, k, d)$ which corresponds to the first structure $(\gamma^{s},\lambda^{n},\gamma^{s})$ and $V_i(d-1, k+1, d)$ that corresponds to the second structure $(\gamma^{s-1},\lambda^{n},\gamma^{s})$.

We can set, in \emph{Lemma 7}, $m=d$, $n=k$ and $s=d$ to calculate $V_i(d, k, d)$, and $m=d-1$, $n=k+1$ and $s=d$ to calculate $V_i(d-1, k+1, d)$. Then we can derive the difference between $V_i(d, k, d)$ and $V_i(d-1, k+1, d)$  as follows.

For $i= 0, \ldots, d-1$, it follows from  \eqref{HI1} that
\begin{align}\label{HID1}
V_i(d, k, d)-V_i(d-1, k+1, d)=\overline{\beta}^{i+1}-\underline{\beta}^{i+1}.
\end{align}

For $i= d, \ldots, k$, we have from \eqref{HI3} and \eqref{HI5}  that
\begin{align}\label{HID2}
&V_i(d, k, d)-V_i(d-1, k+1, d)\nonumber\\
=&\,\,\overline{\beta}^{i-d+1}(\overline{\beta}^d-\alpha^d)-\underline{\beta}^{i-d+1}
(\underline{\beta}^d-\alpha^d).
\end{align}

For $i= k+1$, due to \eqref{HI4} and \eqref{HI5}, there holds
\begin{align}\label{HID3}
&V_i(d, k, d)-V_i(d-1, k+1, d)\nonumber\\
=&\,\,\alpha\overline{\beta}^k(\overline{\beta}-\alpha)+\alpha\overline{\beta}^{k+1}
-\underline{\beta}^{k+2}\nonumber\\
&+\alpha^d(\underline{\beta}^{k+2-d}-\overline{\beta}^{k+2-d}).
\end{align}

For $i= k+2, \ldots, N-1$, due to \eqref{HI4} and \eqref{HI6}, there holds
\begin{align}\label{HID4}
&V_i(d, k, d)-V_i(d-1, k+1, d)\nonumber\\
=&\,\,(i-k)\alpha^{i-k-1}[\alpha\overline{\beta}^k(\overline{\beta}-\alpha)
-\underline{\beta}^{k+1}(\underline{\beta}-\alpha)]
\nonumber\\
&+\alpha^{i-k}(\overline{\beta}^{k+1}-\underline{\beta}^{k+1})
+\alpha^d(\underline{\beta}^{i-d+1}-\overline{\beta}^{i-d+1}).
\end{align}

For $i= N, \ldots, T-1$, due to \eqref{HI2}, there holds
\begin{align}\label{HID5}
&V_i(d, k, d)-V_i(d-1, k+1, d)\nonumber\\
=&\,\,(T-i)\alpha^{i-k-1}[\alpha\overline{\beta}^k(\overline{\beta}-\alpha)
-\underline{\beta}^{k+1}(\underline{\beta}-\alpha)].
\end{align}

Next we focus on the comparison between $V_i(d-1, k, d)$ which corresponds to the second structure $(\gamma^{s-1},\lambda^{n},\gamma^{s})$ and $V_i(d-1, k+1, d-1)$ that corresponds to the first structure $(\gamma^{s},\lambda^{n},\gamma^{s})$. We can set, in \emph{Lemma 7}, $m=d-1$, $n=k$ and $s=d$ to calculate $V_i(d-1, k, d)$, and $m=d-1$, $n=k+1$ and $s=d-1$ to calculate $V_i(d-1, k+1, d-1)$. Then the difference between $V_i(d-1, k, d)$ and $V_i(d-1, k+1, d-1)$ can be given  as follows.\footnote{Here, $k$ should be replaced with $k+1$ since $T=d+k+d$. But in fact, the derivation can proceed if we employ $k$ instead of $k+1$. Hence, for simplicity, $k$ is adopted here.}

For $i= 0, \ldots, d-2$, it follows from \eqref{HI1} that
\begin{align}\label{HID6}
&V_i(d-1, k, d)-V_i(d-1, k+1, d-1)\nonumber\\
=&\overline{\beta}^{i+1}-\underline{\beta}^{i+1}.
\end{align}

For $i= d-1$, due to \eqref{HI1} and \eqref{HI5}, there holds
\begin{align}\label{HID7}
&V_i(d-1, k, d)-V_i(d-1, k+1, d-1)\nonumber\\
=&\,\,\overline{\beta}^d-\underline{\beta}^d+\alpha^{d-1}\overline{\beta}-\alpha^d.
\end{align}

For $i= d, \ldots, k$, we have from \eqref{HI5}  that
\begin{align}\label{HID8}
&V_i(d-1, k, d)-V_i(d-1, k+1, d-1)\nonumber\\
=&\,\,\overline{\beta}^{i-d+1}(\overline{\beta}^d-\alpha^d)
-\underline{\beta}^{i-d+2}(\underline{\beta}^{d-1}-\alpha^{d-1}).
\end{align}

For $i= k+1$, due to \eqref{HI5} and \eqref{HI6}, there holds
\begin{align}\label{HID9}
&V_i(d-1, k, d)-V_i(d-1, k+1, d-1)\nonumber\\
=&\,\,\alpha\overline{\beta}^k(\overline{\beta}-\alpha)+\alpha\overline{\beta}^{k+1}
-\underline{\beta}^{k+2}\nonumber\\
&+\alpha^{d-1}\underline{\beta}^{k+3-d}-\alpha^d\overline{\beta}^{k+2-d}.
\end{align}

For $i= k+2, \ldots, N-1$, due to \eqref{HI6}, there holds
\begin{align}\label{HID10}
&V_i(d-1, k, d)-V_i(d-1, k+1, d-1)\nonumber\\
=&\,\,(i-k)\alpha^{i-k-1}[\alpha\overline{\beta}^k(\overline{\beta}-\alpha)
-\underline{\beta}^{k+1}(\underline{\beta}-\alpha)]
\nonumber\\
&+\alpha^{i-k}(\overline{\beta}^{k+1}-\underline{\beta}^{k+1})
+\alpha^{d-1}\underline{\beta}^{i-d+2}-\alpha^d\overline{\beta}^{i-d+1}.
\end{align}

For $i= N, \ldots, T-1$, due to \eqref{HI2}, there holds
\begin{align}\label{HID11}
&V_i(d-1, k, d)-V_i(d-1, k+1, d-1)\nonumber\\
=&\,\,(T-i)\alpha^{i-k-1}[\alpha\overline{\beta}^k(\overline{\beta}-\alpha)
-\underline{\beta}^{k+1}(\underline{\beta}-\alpha)].
\end{align}

Next we focus on the sign of equations \eqref{HID1}--\eqref{HID11} under the condition 1) in \emph{Lemma 8} that the inequality, $2\alpha\overline{\beta}^{\overline{n}-1}-\underline{\beta}^{\overline{n}}\leq 0$, holds.

In the beginning, it is easy to see that
\begin{align*}
\overline{\beta}^{\overline{n}-n-1}(2\alpha\overline{\beta}^n-\underline{\beta}^{n+1})\leq 2\alpha\overline{\beta}^{\overline{n}-1}-\underline{\beta}^{\overline{n}}\leq 0.
\end{align*}

And we readily have equations \eqref{HID1} and \eqref{HID2} are positive.
According to the condition 1) in \emph{Lemma 8}, there holds
\begin{align*}
&\alpha\overline{\beta}^k(\overline{\beta}-\alpha)+\alpha\overline{\beta}^{k+1}
-\underline{\beta}^{k+2}\\
\leq&\alpha\overline{\beta}^k(\overline{\beta}-\alpha)
-\alpha\overline{\beta}^{k+1}=-\alpha^2\overline{\beta}^k<0,
\end{align*}
which causes that equation \eqref{HID3} is less than 0.

Then  we focus on equations \eqref{HID4} and \eqref{HID5}. Due to the condition, it follows that
\begin{align*}
&\alpha\overline{\beta}^k(\overline{\beta}-\alpha)
-\underline{\beta}^{k+1}(\underline{\beta}-\alpha)\nonumber\\
\leq&\alpha\overline{\beta}^k(\overline{\beta}-\alpha)-\underline{\beta}^{k+2}
+\alpha\overline{\beta}^{k+1}\nonumber\\
\leq&-\alpha^2\overline{\beta}^k<0.
\end{align*}

The above result leads to
\begin{align*}
&(i-k)\alpha^{i-k-1}[\alpha\overline{\beta}^k(\overline{\beta}-\alpha)
-\underline{\beta}^{k+1}(\underline{\beta}-\alpha)]\nonumber\\
&+\alpha^{i-k}(\overline{\beta}^{k+1}-\underline{\beta}^{k+1})\nonumber\\
\leq&\alpha^{i-k-1}[\alpha\overline{\beta}^k(\overline{\beta}-\alpha)
-\underline{\beta}^{k+1}(\underline{\beta}-\alpha)]\nonumber\\
&+\alpha^{i-k}(\overline{\beta}^{k+1}-\underline{\beta}^{k+1})\nonumber\\
=&\alpha^{i-k-1}[2\alpha\overline{\beta}^{k+1}
-\alpha^2\overline{\beta}^k-\underline{\beta}^{k+2}]\leq0,
\end{align*}
from which we readily obtain that equations \eqref{HID4} and \eqref{HID5} are negative.

It is easy to see that equations \eqref{HID6} and \eqref{HID7} are postive.
And from the sign of equation \eqref{HID2}, equation \eqref{HID8} is also greater than 0 since
\begin{align*}
&\overline{\beta}^{i-d+1}(\overline{\beta}^d-\alpha^d)
-\underline{\beta}^{i-d+2}(\underline{\beta}^{d-1}-\alpha^{d-1})\\
>&\underline{\beta}^{i-d+1}(\underline{\beta}^d-\alpha^d)
-\underline{\beta}^{i-d+2}(\underline{\beta}^{d-1}-\alpha^{d-1})\\
=&\underline{\beta}^{i-d+1}\alpha^{d-1}(\underline{\beta}-\alpha)>0.
\end{align*}

And similar to the derivation for equations \eqref{HID3}--\eqref{HID5}, we can prove that equations \eqref{HID9}--\eqref{HID11} are negative.

According to the sign of equations \eqref{HID1}--\eqref{HID11}, we can obtain that $V_i(m,n,s,\overline{\beta},\underline{\beta})$ is a upper concave curve with $n$ for $n>s$. When $n\leq s$, we can similarly derive the expression of $V_i$ and imitate the above proof process to obtain the same result. Therefore, for any attack times $n$ with $\underline{n}\leq n\leq\overline{n}-1$, it follows that $V_i(m,n,s,\theta,\beta)\geq V_i(m,n,s,\overline{\beta},\underline{\beta})\geq \textrm{min} \{V_i(\overline{m},\overline{n}-1,\overline{s},\overline{\beta},\underline{\beta}), V_i(\underline{m},\underline{n},\underline{s},\overline{\beta},\underline{\beta})\}\geq 0$, for $i=0,\ldots,T-1$, where $\overline{m}+\overline{s}=T-\overline{n}+1$, $\overline{m}=\overline{s}$ or $\overline{m}=\overline{s}-1$, $\underline{m}+\underline{s}=T-\underline{n}$, and $\underline{m}=\underline{s}$ or $\underline{m}=\underline{s}-1$. The proof is completed.
\end{IEEEproof}

Finally, \emph{Theorem 3} is proved as follows.

\begin{IEEEproof}[Proof of Theorem 3]
Let $\mathbb{N}=\{0,1,2,\ldots\}$ and $q(j|s,a)=\sum_{i=j}^\infty Pr(h^i(\overline{P})|s,a)$. According to Theorem 4.7.4 in \cite{Puterman2005}, it suffices to prove that the following items are true for $k=1,\ldots,T$.
\begin{enumerate}
  \item $R_k(s,a)$ is nondecreasing in $s$ for all $a\in \mathbb{\tilde{A}}$;
  \item $q(j|s,a)$ is nondecreasing in $s$ for all $j\in \mathbb{N}$ and $a\in \mathbb{\tilde{A}}$;
  \item $R_k(s,a)$ is a superadditive function on $\mathbb{\tilde{S}}\times \mathbb{\tilde{A}}$;
  \item $q(j|s,a)$ is a superadditive function on $\mathbb{\tilde{S}}\times \mathbb{\tilde{A}}$ for all $j\in \mathbb{N}$;
  \item $R_{T+1}(s)$ is nondecreasing in $s$.
\end{enumerate}

Take $R_k$ which corresponds to the average error for example. For a given $a$, $R_k(s,a)=\textrm{Tr}((1-\beta_k(a))\overline{P}+\beta_k(a)h(s))-\omega a$ is nondecreasing in $s$ due to \emph{Lemma 2}, from which the proof of 1) is completed.

Then for $s^+\geq s^-$ in $\mathbb{\tilde{S}}$ and $a^+\geq a^-$ in $\mathbb{\tilde{A}}$, there hold
$
R_k(s^+,a^+)-R_k(s^-,a^+)=\textrm{Tr}(\beta_k(a^+)[h(s^+)-h(s^-)])
$
and
$
R_k(s^+,a^-)-R_k(s^-,a^-)=\textrm{Tr}(\beta_k(a^-)[h(s^+)-h(s^-)]),
$
which, in light of \eqref{pk}--\eqref{pksinr} and \emph{Lemma 2}, lead to
\begin{align*}
R_k(s^+,a^+)-R_k(s^-,a^+)\geq R_k(s^+,a^-)-R_k(s^-,a^-).
\end{align*}
And thereby, from \emph{Definition 3}, the proof of 3) is completed.

To prove 2) and 4), the expression of $q(j|s,a)$ is presented based on \eqref{Prt} as follows.
For $j=0$, $q$ is given by $q(0|s,a)=1$, and for $j=1,2,\ldots$ and a given $a$, it follows that
\begin{align*}
q(j|s,a)= \left\{ \begin{array}{ll}
 0,& \textrm{if}\, s<h^{j-1}(\overline{P}),\\
\beta(a),& \textrm{if}\, s\geq h^{j-1}(\overline{P}),
\end{array} \right.
\end{align*}
from which 2) is obtained.

For $j=0$, 4) holds obviously. For $j=1,2,\ldots$, there are three cases. First, suppose that $s^+,s^-<h^{j-1}(\overline{P})$. Apparently 4) holds. Second, suppose that $s^-<h^{j-1}(\overline{P})$ and $s^+\geq h^{j-1}(\overline{P})$. This generates $q(j|s^-,\cdot)=0$ and $q(j|s^+,a)=\beta(a)$. Then we obtain 4) since $\beta(a^+)\geq \beta(a-)$. Third, suppose that $s^+,s^-\geq h^{j-1}(\overline{P})$. Then we have $q(j|s^+,a)=q(j|s^-,a)$ which implies that 4) is true.

5) is a direct result from the fact that $R_{T+1}=0$. Then the proof is completed.
\end{IEEEproof}
\end{appendix}

\bibliographystyle{IEEE}

\begin{thebibliography}{10}
\bibitem{Johansson2014}K. H. Johansson, G. J. Pappas, P. Tabuada and C. J. Tomlin, ``Guest editorial special issue on control of cyber-physical systems," \emph{IEEE Trans. Autom. Control}, vol. 59, no. 12, pp. 3120--3121, 2014.
\bibitem{Poovendran2012}R. Poovendran, K. Sampigethaya, S. K. S. Gupta, I. Lee, K. V. Prasad, D. Corman, and J. Paunicka, ``Special issue on cyber-physical systems," \emph{Proc. IEEE}, vol. 100, no. 1, pp. 6--12, 2012.
\bibitem{Fawzi2014}H. Fawzi, P. Tabuada, and S. Diggavi, ``Secure estimation and control for cyber-physical systems under adversarial attacks," \emph{IEEE Trans. Autom. Control}, vol. 59, no. 6, pp. 1454--1467, 2014.
\bibitem{Sundaram2011}S. Sundaram and C. N. Hadjicostis, ``Distributed function calculation via linear iterative strategies in the presence of malicious agents," \emph{IEEE Trans. Autom. Control}, vol. 56, no. 7, pp. 1495--1508, 2011.
\bibitem{Pajic2014}M. Pajic, J. Weimer, N. Bezzo, P. Tabuada, O. Sokolsky, I. Lee, and G. J. Pappas, ``Robustness of attack-resilient state estimators," in \emph{Proc. International Conference of Cyberphysical Systems (ICCPS)}, 2014, pp. 163--174.
\bibitem{Mo2015} Y. Mo and B. Sinopoli, ``Secure estimation in the presence of integrity attacks," \emph{IEEE Trans. Autom. Control}, vol. 60, no. 4, pp. 1145--1151, 2015.
\bibitem{Vamvoudakis2014} K. G. Vamvoudakis and J. P. Hespanha, B. Sinopoli, and Y. Mo, ``Detection in adversarial environments," \emph{IEEE Trans. Autom. Control}, vol. 59, no. 12, pp. 3209--3223, 2014.
\bibitem{Brown2017} H. E. Brown and C. L. DeMarco, ``Risk of cyber-physical attack via load with emulated inertia control," \emph{IEEE Trans. Smart Grid}, doi: 10.1109/TSG.2017.2697823.
\bibitem{Amini2018} S, Amini, F. Pasqualetti, and H. Mohsenian-Rad, ``Dynamic Load Altering Attacks Against Power System Stability: Attack Models and Protection Schemes," \emph{IEEE Trans. Smart Grid}, vol. 9, no. 4, pp. 2862--2872, 2018.
\bibitem{Jiang2018} F. Jiang, Y. Fu, B. B. Gupta, F. Lou, S. Rho, F. Meng, and Z. Tian, ``Deep Learning based Multi-channel intelligent attack detection for Data Security," \emph{IEEE Trans. Sustainable Computing}, doi: 10.1109/TSUSC.2018.2793284.
\bibitem{Amin2013a} S. Amin, X. Litrico, S. Sastry, and A. M. Bayen, ``Cyber security of water SCADA systems--Part I: Analysis and experimentation of stealthy deception attacks," in \emph{IEEE Trans. Control Syst. Technol.}, vol. 21, no. 5, pp. 1963--1970, 2013.
\bibitem{Amin2013b} S. Amin, X. Litrico, S. Sastry, and A. M. Bayen, ``Cyber security of water SCADA systems--Part II: Attack detection using enhanced hydrodynamic models," in \emph{IEEE Trans. Control Syst. Technol.}, vol. 21, no. 5, pp. 1679--1693, 2013.
\bibitem{Mo2009} Y. Mo and B. Sinopoli, ``Secure control against replay attacks," in \emph{47th Annual Allerton Conference on Communication, Control, and Computing}, 2009, pp. 911--918.
\bibitem{Zhu2014} M. Zhu and S. Mart\'{i}nez, ``On the performance analysis of resilient networked control systems under replay attacks," \emph{IEEE Trans. Autom. Control}, vol. 59, no. 3, pp. 804--808, 2014.
\bibitem{Mo2014} Y. Mo, R. Chabukswar, and B. Sinopoli, ``Detecting integrity attacks on SCADA systems," \emph{IEEE Trans. Control Syst. Technol.}, vol. 22, no. 4, pp. 1396--1407, 2014.
\bibitem{Befekadu2015}G. Befekadu, V. Gupta, and P. Antsaklis, ``Risk-sensitive control under markov modulated denial-of-service (DoS) attack strategies," \emph{IEEE Trans. Autom. Control}, vol. 60, no. 12, pp. 3299--3304, 2015.
\bibitem{Peris2015} C. D. Peris and P. Tesi, ``Input-to-state stabilizing control under denial-of-service," \emph{IEEE Trans. Autom. Control}, vol. 60, no. 11, pp. 2930--2944, 2015.
\bibitem{Foroush2012}H. S. Foroush and S. Mart\'{i}nez, ``On event-triggered control of linear systems under periodic denial of service attacks," in \emph{Proc. IEEE Conf. Decision Control}, Maui, HI, USA, 2012, pp. 2551--2556.
\bibitem{Poisel2011} R. Poisel, \emph{Modern Communications Jamming: Principles and Techniques.} Artech House, 2011.
\bibitem{Zhang2015}H. Zhang, P. Cheng, L. Shi, and J. Chen, ``Optimal denial-of-service attack scheduling with energy constraint," in \emph{IEEE Trans. Autom. Control}, vol. 60, no. 11, pp. 3023--3028, 2015.
\bibitem{Zhang2016}H. Zhang, P. Cheng, L. Shi, and J. Chen, ``Optimal DoS attack scheduling in wireless networked control system," \emph{IEEE Trans. Control Syst. Technol.}, vol. 24, no. 3, pp. 843--852, 2016.
\bibitem{Li2015b}Y. Li, L. Shi, P. Cheng, J. Chen, and D. E. Quevedo, ``Jamming attacks on remote state estimation in cyber-physical systems: A game-theoretic approach," \emph{IEEE Trans. Autom. Control}, vol. 60, no. 10, pp. 2831--2836, 2015.
\bibitem{Zhang2018}H. Zhang, Y. Qi, J. Wu, L. Fu, and L. He, ``DoS attack energy management against remote state estimation," \emph{IEEE Trans. Control Netw. Syst.}, vol. 5, no. 1, pp. 383--394, 2018.
\bibitem{Zhang2017}H. Zhang, Y. Qi, and J. Wu, ``Optimal jamming power allocation against remote state estimation", in \emph{Proc. IEEE American Control Conference}, 2017, pp. 2378--5861.
\bibitem{Qin2018} J. Qin, M. Li, L. Shi, and X. Yu, ``Optimal Denial-of-Service Attack Scheduling with Energy Constraint Over Packet-dropping Networks," \emph{IEEE Trans. Autom. Control}, vol. 63, no. 6, pp. 1648--1663, 2018.
\bibitem{Shi2012} L. Shi and L. Xie, ``Optimal sensor power scheduling for state estimation of gauss-markov systems over a packet-dropping network," \emph{IEEE Trans. Signal Processing,} vol. 60, no. 5, pp. 2701--2705, 2012.
\bibitem{Savage2009} C. O. Savage and B. F. La Scala, ``Optimal scheduling of scalar Gauss--Markov systems with a terminal cost function," \emph{IEEE Trans. Autom. Control}, vol. 54, no. 5, pp. 1100--1105, 2009.
\bibitem{Horn2012}R. A. Horn and C. R. Johnson, \emph{Matrix Analysis}. Cambridge, U.K.: Cambridge University Press, 2012.
\bibitem{Shaked2007}M. Shaked, J. G. Shanthikumar, \emph{Stochastic Orders}. New York, NY, USA: Springer--Verlag, 2007.
\bibitem{Peng2017} L. Peng, L. Shi, X. Cao, and C. Sun, ``Optimal Attack Energy Allocation against Remote State Estimation," \emph{IEEE Trans. Autom. Control}, doi: 10.1109/TAC.2017.2775344.
\bibitem{Puterman2005} M. L. Puterman, \emph{Markov decision processes: discrete stochastic dynamic programming}. John Wiley \& Sons, 2005.
\bibitem{Li2017}Y. Li, D. E. Quevedo, S. Dey, and L. Shi, ``SINR-based DoS attack on remote state estimation: A game-theoretic approach," \emph{IEEE Trans. Control Netw. Syst.}, vol. 4, no. 3, pp. 632--642, 2017.
\bibitem{Adibi2007} M. Adibi and V. T. Vakili, ``Comparison of cooperative and non-cooperative game schemes for SINR-constrained power allocation in multiple antenna cdma communication systems," in \emph{Proc. IEEE Int. Conf. Signal Process. Commun.}, 2007, pp. 1151--1154.
\end{thebibliography}

\end{document}